\documentclass[12pt]{JHEP3}
\usepackage[dvips]{epsfig}
\usepackage{epsfig}
\usepackage{graphicx}

\title{Braneworld Stars and Black Holes }
\author{~
Simon Creek, Ruth Gregory, Panagiota Kanti, and Bina Mistry\\
Centre for Particle Theory, Department of Mathematical Sciences,\\
Durham University, South Road, Durham, DH1 3LE, U.K.}

\abstract{
We look for spherically symmetric star or black hole
solutions on a Randall-Sundrum brane from the perspective of the bulk.
We take a known bulk solution, and analyse possible braneworld
trajectories within it that correspond, from the
braneworld point of view, to solutions of the brane 
Tolman-Oppenheimer-Volkoff equations. Our solutions are therefore
embedded consistently
into a full bulk solution. We find the full set of static gravitating
matter sources on a brane in a range of bulk spacetimes, analyzing
which can correspond to physically sensible sources. Finally, we
look at time-dependent trajectories in a Schwarzschild--anti de
Sitter spacetime as possible descriptions of time-dependent braneworld
black holes, highlighting some of the general features one might
expect, as well as some of the difficulties involved in getting
a full solution to the question.
}
\preprint{hep-th/0606006\\
DCPT-06/13}

\def\cee{{\relax\hbox{$\inbar\kern-.3em{\rm C}$}}}

\newcommand{\be}{\begin{equation}}
\newcommand{\ee}{\end{equation}}
\newcommand{\bea}{\begin{eqnarray}}
\newcommand{\eea}{\end{eqnarray}}

\begin{document}

\section{Introduction / Motivation}

The idea that spacetime may not be simply four-dimensional, but have
extra dimensions as yet undetected by experiment, has become essentially
accepted as fact over the past two decades, largely as a consequence
of string or M-theory, but also as a result of earlier work on
supergravity. 
Over the past few years, a new alternative has emerged in our understanding
of how extra dimensions can be compactified -- braneworlds and warped
compactifications. Rather than the old Kaluza-Klein (KK) idea of wrapping
up extra dimensions so small we only see them through extra massless
fields, the braneworld idea allows us to have relatively large extra
dimensions, possibly even up to the microscale, with standard particle
physics confined to the ``brane'' and thereby unable to detect the
extra dimensions at ordinary energy scales \cite{EBW,ADD,RS}.
Gravity however can sample these extra dimensions, and one of the most
alluring aspects of warped compactifications is the possibility of 
unusual gravitational phenomenology not only at small scales (akin
to the KK picture) but at large scales, too \cite{oxmill,GRS,DGP}.

The braneworld paradigm 
views our universe as a slice of some higher dimensional spacetime, 
in which we have standard four-dimensional physics confined 
to the brane, and only gravity (plus possibly
a small number of other fields) propagating in the bulk.
Confinement to the brane, while at first sounding counter-intuitive,
is in fact a common occurrence. The first braneworld scenarios
\cite{EBW} used topological defects to model the braneworld, and
zero-modes on the defects to produce confinement, and in 
string theory, D-branes have `confined' gauge theories on their
worldvolumes.  The new
phenomenology of braneworld scenarios is then primarily located
in the gravitational sector, with a particularly nice possible
resolution of the hierarchy problem being its primary motivation.
Clearly however, the scenario has far outgrown these 
initial particle phenomenology
motivations, and has proved a fertile testbed for new possibilities
in cosmology, astrophysics, and quantum gravity.
One of the most popular models to
explore, and the one which we will be using, has been the 
Randall-Sundrum scenario, \cite{RS}, which
consists of a domain wall universe living in five-dimensional
anti-de Sitter (adS) spacetime. This model can be loosely motivated
by the Horava-Witten compactification of M-theory \cite{HW}, and many
of the ideas tested and developed in this simple, calculationally explicit
model underly more recent string theory motivated compactifications
\cite{KKLT}.

The Randall-Sundrum model has one (or two) domain walls situated
as minimal submanifolds in adS spacetime. In its usual form, 
the metric of the braneworld is
\be
ds^2 = e^{-2k|z|} \left [ -dt^2 + d{\bf x}^2 \right ] + dz^2\,.
\label{rsmet}
\ee
Here, the spacetime is constructed so that there are four-dimensional
flat slices stacked along the fifth $z$-dimension, which have a 
$z$-dependent conformal pre-factor known as the warp factor. Since
this warp factor has a cusp at $z=0$, this indicates the presence of a
domain wall -- the braneworld -- which represents an exactly flat 
Minkowski universe. The reason for choosing this particular slicing
of adS spacetime was to have a flat Minkowski metric on the brane --
i.e.\ to choose the ``standard vacuum''.

Randall and Sundrum showed that although 
gravity was inherently five-dimen\-sio\-nal, and the spacetime was
strongly warped, as far as a four-dimensional brane-world observer was
concerned, the gravitational potential of a particle on the brane was
the Newtonian $1/r$ potential to leading order. 
A complete analysis shows that the 
graviton propagator has the correct tensor structure, 
and that the effect of the KK modes is to introduce a $1/r^3$
correction to the gravitational potential \cite{GT}.

In astrophysics and cosmology we are not so much
interested in the perturbative graviton propagator as in 
issues such as cosmological models or black holes, which are
questions of non-perturbative, or strong gravity.
The braneworld generalization of the FRW universe has been 
well explored and understood \cite{GCOS}; the high
degree of symmetry present renders the full five-dimensional
problem fully integrable \cite{BCG}, and the general cosmological
braneworld is fully understood in terms of a slice of a 
five-dimensional adS black hole \cite{BCOS}. The mass of this 
bulk black hole then generates a radiation-style source for the Friedman
equation. 
Interestingly, this understanding feeds in to the second question:
{\it What is the metric of a braneworld black hole?}

At first sight it might seem that this question should be 
very similar to answer; as both reduce to a two-dimensional problem,
however, the symmetry groups of the two spacetimes
are crucially different. For cosmology, the metric splits into two
parts -- the two dimensions on which it depends, and the 
spatial part of the universe, which has constant curvature.
This problem is equivalent to 
a two-dimensional field theory which turns out to be totally
integrable. For the black hole however, the metric splits into 
three parts -- the two dimensions on which it depends, the time 
coordinate and the remaining spatial part in which the horizon 
resides. Thus there are two fields in the two-dimensional theory, 
and there is no longer a simple solution \cite{CG}.

The first attempt, \cite{CHR}, 
to find a black hole solution replaced the Minkowski metric 
in (\ref{rsmet}) by the Schwarzschild metric, thus creating a 
black string sticking out of the brane.  Unfortunately, as 
suspected by the authors, this string is unstable to classical 
linear perturbations \cite{BSINS}. Chamblin et.\ al.\ realised
that the true static localised black hole would actually be a slice of a 
five-dimensional accelerating black hole metric (known as 
the C-metric, \cite{CMET}, in four dimensions), however no
such metric has as yet been found. A lower dimensional
version of a black hole living on a $2+1$-dimensional braneworld
was however presented by Emparan, Horowitz and Myers \cite{EHM},
using this four-dimensional C-metric. Since then, several authors
have attempted to find the full metric using numerical techniques
\cite{BHNUM}, although the main drawback seems to be that it is
a very sensitive numerical system. Nonetheless, the results of
\cite{KU} for small black holes are encouraging.

Analytically, progress has mostly (though see \cite{bulkbh})
been limited to considering the brane metric equations of motion, 
with the only bulk input coming from the projection of the Weyl 
tensor, the ``{\it Weyl term}'' \cite{SMS}, onto the brane.
Since this system contains an unknown bulk dependent term, 
assumptions have to be made either in the form of the metric
or the Weyl term \cite{BBH,GWBD}. There is no
clear consensus on what the brane black hole metric is, however, some
interesting features which do occur are wormholes and singular
horizons \cite{IFeat,GWBD}.

One of the reasons this braneworld black hole metric is so interesting
is that it is believed to correspond to a quantum corrected black hole.
In string theory, it has been realized for some time that there
is a correspondence between string theory on adS space, and a
CFT on the boundary of that adS space \cite{MAL}. In other 
words, all of the information contained in the five-dimensional 
gravitational spacetime is encoded in a pure quantum field theory 
living on a four-dimensional spacetime. 
In the braneworld picture, the brane is not at the adS boundary, 
but at a finite distance, and the theory on this brane now 
contains gravity, as well as a conformal energy-momentum tensor 
-- the Weyl term. The effect of the brane 
on the adS/CFT correspondence therefore is that the bulk theory 
of gravity in five dimensions corresponds to the four dimensional 
brane theory of a CFT, with a UV cutoff, interacting with gravity 
\cite{DL}.  Since the brane theory is a quantum 
theory, the holographic correspondence suggests that the classical
bulk solution projects to a quantum corrected solution on the
brane \cite{EFK}. Indeed, it was this type of argument that led Tanaka
\cite{TAN} to argue that the braneworld black hole metric would be
time-dependent, corresponding to the back-reaction of Hawking radiation
on the black hole metric.

For cosmological solutions, this holographic interpretation works 
very well; the presence of a black hole horizon in the bulk (which is
the only allowed class of bulk solution \cite{BCG}) induces a corresponding
source in the Friedman equation which has the form of a radiation source.
This source can be interpreted as a CFT in a thermal
state corresponding to the Hawking temperature of the bulk black hole.
The brane cosmological metric has a constant curvature spatial part,
and its symmetries demand that only a radiation energy Weyl term is
allowed. From the bulk perspective, this means that every point on
the brane is at the same distance from the bulk black hole. Thus a 
flat universe corresponds to a `flat' bulk black hole, a closed 
universe to a conventional spherical bulk black hole. 

Transporting this intuition over to the brane black hole situation,
one can imagine that the black hole becomes displaced from the 
``center of gravity'' of the spherical wall, causing an anisotropy in
the brane Weyl term, ${\cal E}_{\mu\nu}$. As the black
hole gets closer to the brane, this anisotropy increases, possibly becoming
more important than the radiation term. This reasoning argues for a
near-horizon equation of state for ${\cal E}_{\mu\nu}$ which leads to
a singular `event horizon' \cite{GWBD}, which possibly corresponds to 
a Boulware choice of vacuum in the quantum corrected black hole \cite{EFK}.

However, there is another holographic interpretation possible, and
one which is far more intriguing and experimentally relevant -- one which
incorporates black hole evaporation.
Instead of imagining a quasi-static transport of the 
bulk black hole towards the brane, consider a bulk geodesic. From the 
point of view of the brane, these trajectories have constant acceleration
away from the brane, so a particle in the bulk moving along a geodesic
initially moves towards the brane, can touch the brane, but then moves back
into the bulk accelerating away (see section \ref{sec:tdep}). 
Thus a black hole would
move towards the brane, hit the brane, then recoil away back off to
infinity. From the brane point of view, this would correspond to collapse
of conformal matter, localized around the lightcone, formation of an
horizon, and subsequent evaporation of the black hole, again localized
around the lightcone. This picture was indeed obtained in perturbation
theory in \cite{grs}, where the metric of a particle leaving the brane
was obtained to leading order.

Such a picture should be a reasonable approximation for small mass black 
holes, which, coincidentally, are precisely the type of black holes that
are believed to be important in LHC and cosmic ray phenomenology
(for some early works, see \cite{BHC,CRS}; for a more complete list
of references, see \cite{Kanti}). Such small black holes are thought to
be produced after brane-localized particles scatter at high energies and
undergo gravitational collapse. A horizon is then formed engulfing the
two particles, which can never escape their mutual gravitational attraction.
Due to their small mass, these black holes quickly evaporate through the
emission of Hawking radiation \cite{Kanti}: for a black hole with mass
$M_{BH} \simeq 5$ TeV, and fundamental gravity scale $M_*=1$ TeV,
their lifetime is only $\tau_{BH} \simeq 10^{-26}$ sec, and therefore
exist on our brane only momentarily. Even for a higher mass, that would
result in a longer lifetime, the corresponding black hole may still
`disappear' from our brane due to the so-called recoil effect
\cite{Frolov-recoil, RECOIL, Stojkovic}. Due to the absence of an
analytic solution describing a black hole localized on a brane with
a non-vanishing self-energy, all studies of the evaporation of
brane-world black holes have been restricted to the case where the
black hole mass is assumed to be significantly larger than the
brane self-energy. In addition, by assuming that the black hole
horizon is much smaller than the inverse adS radius, the bulk
warping has also been ignored. As a result, all studies up to now
have failed to consider the complete bulk-brane-black-hole
gravitational system. 

In the present work, we study the aforementioned
gravitational system in full. Our analysis will be complementary to
work on probe branes \cite{Flachi}, and develops the work on specific
brane trajectories in black hole backgrounds \cite{Seahra,Galfard}.
We restrict our study to the case of a 5-dimensional spacetime
in which a 3-brane with a non-negligible energy-momentum tensor is
embedded. By using the Israel's junction conditions \cite{Israel},
we derive a set of equations corresponding to a spherically
symmetric brane with additional matter content corresponding
to a homogeneous and isotropic fluid, in other words the brane
equivalent of the Tolman-Oppenheimer-Volkoff (TOV) equations. The main
difference between this work and the brane based work of \cite{BBH}
is that we have not only a complete brane solution to the TOV
equations, but also the full bulk solution. In other words a genuine
brane star. Clearly the general brane-bulk system has infinitely many
degrees of freedom, so our approach here is to restrict to a spherically
symmetric bulk solution, and a variety of bulk backgrounds are 
considered with the final objective being the consistent embedding
of a 3-brane into a Schwarzschild--anti de Sitter spacetime.

The outline of the paper is as follows: In the next section we
derive the brane equations of motion for a brane with a general
isotropic fluid source living in a (general) spherically symmetric
bulk. We then consider the static system in section \ref{sec:genstat},
and show how the static brane is completely integrable. In section
\ref{sec:star} we specialize to the physically relevant case of
a Schwarzschild--anti de Sitter bulk, exploring possible black hole
and stellar solutions. We then briefly consider time dependent
solutions in section \ref{sec:tdep} before concluding.

%%%%%%%%%%%%%%%%%%%%%%%%%%%%%%%%%%%%%%%%%%%%%%%%%%%%%%%%%%%%%%%%%%%%%%

\section{The general brane equations}

We now adopt a spherically-symmetric coordinate system, and write the
5-dimensio\-nal line-element in the form
\begin{eqnarray} \label{eq:metric}
ds^2 &=& -U(r)\,d\tau^2 + \frac{1}{U(r)}\,dr^2 +
r^2(d\chi^2 + \sin^2\chi\,d\Omega_{_{\rm{II}}}^2)\,, \label{general}
\end{eqnarray}
where $U(r)$ is a general function of the global radial coordinate and
$d\Omega_{_{\rm{II}}}^2$ the line element on a unit 2-sphere.
We are considering configurations within the second Randall-Sundrum
model with a single brane of positive tension and
with spacetime reflection symmetric in the wall.

The location of the brane in the 5-dimensional bulk is described by
the function $\chi(\tau, r)$. If we take a 5-vector to be described by
the component functions $x^\mu = (\tau,r,\chi,\theta,\phi)$, we may
form a new basis in terms of the (unnormalised) tangent vectors
and the unit normal:
\begin{eqnarray}
T^\mu &=& (1,0,\dot \chi,0,0) \nonumber \\
R^\mu &=& (0,1,\chi^\prime,0,0) \nonumber \\
\Theta^\mu &=& (0,0,0,1,0) \label{basis} \\
\Phi^\mu &=& (0,0,0,0,1) \nonumber \\
n_\mu &=& n(- \dot \chi, -\chi^\prime,1,0,0)\,. \nonumber
\end{eqnarray}
In the above, overdot and prime denote partial differentiation
with respect to $\tau$ and $r$, respectively, and
$\frac{1}{n^2} = \left(-\frac{\dot\chi^2}{U} + U{\chi^\prime}^2 +
\frac{1}{r^2}\right)$. The tensor $h_{\mu\nu} = g_{\mu\nu}
- n_\mu n_\nu$ projects vectors onto the wall, and its tangential
components define the induced metric on the
brane. In the aforementioned basis, it can be simply evaluated as
\begin{displaymath}
h_{\mu\nu} =
\left( \begin{array}{ccccc}
-U + r^2 \dot \chi^2 & r^2\dot \chi \chi^\prime  & & & \\
r^2\dot \chi \chi^\prime & \frac{1}{U} + r^2{\chi^\prime}^2 & & & \\
&  & r^2\sin^2\chi & & \\
& & & r^2\sin^2\chi \sin^2\theta & \\
& & & & 0\\
\end{array} \right)\,.
\end{displaymath}

The Israel junction conditions \cite{Israel}, triggered by the presence 
of the brane with a non-vanishing distributional
energy-momentum tensor $T_{\mu\nu}$, take the form
\begin{eqnarray}
\left[K_{\mu\nu} - Kh_{\mu\nu}\right]^+_- = \kappa_5 T_{\mu\nu}\,,
\end{eqnarray}
where $K_{\mu\nu} = h^\rho_\mu h^\sigma_\nu \nabla_\rho n_\sigma$
is the extrinsic curvature of the brane, and $\kappa_5 = 8\pi G_5$.
Using the $Z_2$ reflection symmetry around the wall, the
Israel conditions can be rewritten as
\begin{eqnarray}
K_{\mu\nu}  = \frac{\kappa_5}{6} \left(3T_{\mu\nu} - h_{\mu\nu}\,T\right)\,.
\end{eqnarray}

Deviating from the simplified ansatz of \cite{CHR} in which the brane
was characterized only by a constant self-energy, in this work we assume
that the energy-momentum tensor on the brane may take the general form
\begin{eqnarray}
T_{\mu\nu} = \left[\rho (\tau,r) + p(\tau,r)\right]h_{\mu\sigma}
h_{\nu\rho}\,u^\sigma u^\rho + p(\tau,r)\,h_{\mu\nu}\,.
\end{eqnarray}
The above ansatz describes an isotropic distribution of a perfect fluid
with
$\rho (\tau,r)$ and $p(\tau,r)$ the energy density and pressure,
respectively, of the
fluid. The vector $u^\mu$ is the fluid's 4-velocity, that satisfies the
normalization
condition $u^\mu u^\nu h_{\mu\nu} = -1$. It can be easily shown that
working in the
rest frame of the fluid on the brane, i.e. writing
%%%%%%%%%%%%
\begin{equation}
u^\mu= \frac{1}{\sqrt{-h_{TT}}}\,(1,0,0,0,0)
\label{velocity}
\end{equation}
%%%%%%%%%%%%
in the basis ($T,R,\Theta,\Phi,n$), is equivalent to taking $u^\mu$ to
be parallel to
the time-like tangent vector $T^\mu$ in the original basis ($\tau, r,
\chi, \theta, \phi$).
The ansatz (\ref{velocity}) allows us to rewrite the brane
energy-momentum tensor as:
\begin{equation}\label{energy-mom}
T_{\mu\nu} = - (\rho + p)\,\frac{h_{\mu T}h_{\nu T}}{h_{TT}} +
p\,h_{\mu\nu}\,.
\end{equation}
For convenience, we will write the energy density and pressure of the perfect
fluid on the brane
in terms of an ``equation of state'' $p(\tau,r) = w(\tau,r) \rho
(\tau,r)$, and define for
later convenience the quantity
%%%%%%%%%%%%%
\begin{equation}
v(\tau,r) \equiv 2 + 3 w(\tau,r)\,.
\label{v}\end{equation}
%%%%%%%%%%%%%%%
Note however, that unlike a standard equation of state, in which the
pressure is a fixed multiple of the energy, this redefinition does not
restrict the pressure in any way since $v$ is an arbitrary function of
both $r$ and $\tau$.

By using the form of the energy momentum tensor given in (\ref{energy-mom})
together with (\ref{v}), the Israel conditions for a brane
containing a perfect fluid take the form
\begin{equation}
   K_{\mu\nu} = \frac{\kappa_5}{6}\,\rho \left[ h_{\mu\nu} -
   (1+v)\,\frac{h_{\mu T}h_{\nu T}}{h_{TT}}\right], \label{Israel2}
\end{equation}
or, more explicitly,
  \begin{eqnarray}
&~& \hspace*{-1.1cm} K_{TT} = - n\left(\ddot \chi + Ur\chi^\prime
\dot\chi^2 -
    \frac{1}{2}\,U U^\prime \chi^\prime\right) =
    - \frac{\kappa_5}{6}\,\rho v \left(-U + r^2 \dot \chi^2\right)
\label{TT} \\[1mm]
&~& \hspace*{-1.1cm} K_{RR} = - n\left(\chi^{\prime\prime} +
\frac{2\chi^\prime}{r} +
    \frac{U^\prime \chi^\prime}{2U} + Ur {\chi^\prime}^3\right)
%\nonumber\\
    = \frac{\kappa_5}{6} \rho\left[\frac{1}{U} + r^2{\chi^\prime}^2 +
     \frac{ (1+v)\,r^4\dot \chi^2 {\chi^\prime}^2}{U - r^2 \dot
\chi^2}\right]
\label{RR}\\[1mm]
&~& \hspace*{-1.1cm} K_{TR} = - n\left(\dot \chi^\prime + \frac{\dot
\chi}{r} +
    U r {\chi^\prime}^2\dot\chi - \frac{U^\prime \dot \chi}{2U}\right) =
     - \frac{\kappa_5}{6}\,\rho v r^2\dot \chi \chi^\prime
    \label{TR} \\[1mm]
&~& \hspace*{-1.1cm} K_{\Theta\Theta} = - n \left(Ur\chi^\prime
\sin^2\chi -
    \sin\chi \cos\chi\right) = \frac{\kappa_5}{6}\,\rho r^2\sin^2\chi\,.
\label{ThTh}
    \end{eqnarray}
%%%%%%%%%%%%%%
Two additional, but not independent, equations are obtained
from conservation of energy-momentum:
\begin{eqnarray}
\dot\chi \left(1+\frac{1}{2}U^\prime r - U + \frac{\kappa_5\rho}{6n}
(1+v)r^2
\cot\chi\right) &=& - \frac{\kappa_5 \dot \rho}{6n}r^2\,,
\label{con1}\\[1mm]
\chi^\prime \left(1+\frac{1}{2}U^\prime r - U +
\frac{\kappa_5\rho}{6n}\,(1+v)
\frac{r^4\dot\chi^2\cot\chi}{-U + r^2 \dot \chi^2}\right) &=& -
\frac{\kappa_5\rho^\prime}{6n}r^2\,. \label{con2}
\end{eqnarray}

We may summarise the above results by saying that a 4-dimensional brane
containing a perfect fluid, described by its energy density $\rho(\tau,r)$
and equation of state $p(\tau,r) = w(\tau,r)\rho(\tau,r)$, can be
successfully embedded in a 5-dimensional spherically-symmetric background
defined by a single function $U(r)$ as long as we can find a consistent
set of functions $\rho(\tau,r)$, $w(\tau,r)$ and $\chi(\tau,r)$
-- with the latter denoting the position of the brane in the
5-dimensional spacetime -- satisfying (\ref{TT})-(\ref{con2}).
This task is simplified if we define a new function $\alpha = r\cos\chi$,
in terms of which equations (\ref{TT})-(\ref{con2}) may be written as
%%%%%%%%%%%%%%%%%%
\begin{eqnarray}
&~& \hspace*{-0.3cm} \frac{r^2 \ddot{\alpha}}{U} - (\alpha^\prime r -
\alpha)
\left(\frac{U'r}{2} - U\right)+ \alpha + \nonumber \\
&~&\hspace*{3.5cm}\frac{(1+v)}{U}
\left[U(\alpha^\prime r - \alpha) + \alpha\right]
\left(-U + \frac{r^2{\dot\alpha}^2}{r^2-\alpha^2}\right)=0\,,
\label{eq:IsraelTTalpha}\\[1mm]
&~& \hspace*{-0.3cm} Ur^2 \alpha^{\prime\prime} + (\alpha^\prime r -
\alpha)
\left(\frac{U'r}{2} - U\right) - \alpha + \nonumber \\
&~&\hspace*{2.8cm} (1+v)\frac{\left[U(\alpha^\prime r - \alpha) +
\alpha\right]Ur^2{\dot\alpha}^2 (\alpha^\prime r - \alpha)^2}
{(r^2-\alpha^2)\left(r^2{\dot\alpha}^2-U({r^2-\alpha^2})\right)} 
= 0\,,  \label{eq:IsraelRRalpha} \\[1mm]
&~& \hspace*{1.8cm}  r \dot \alpha^\prime - \frac{1}{2}\frac{U^\prime
r}{U} \dot \alpha
+   (1+v)(U(\alpha^\prime r - \alpha) +
\alpha)\frac{\dot\alpha(\alpha^\prime r -
\alpha)}{(r^2-\alpha^2)} = 0\,,\label{eq:IsraelTRalpha}\\[1mm]
&~& \hspace*{1.7cm} U(\alpha^\prime r - \alpha ) + \alpha 
= \frac{\kappa_5}{6}\rho r
\left(-\frac{r^2{\dot \alpha}^2}{U} + U(\alpha^\prime r - \alpha )^2 +r^2 -
\alpha^2\right)^{\frac{1}{2}},  \label{eq:Israelthetathetaalpha}\\[1mm]
&~& \hspace*{4.1cm}  \frac{\dot\rho r}{\rho} = r\dot\alpha \left(
\frac{1 +
\frac{1}{2}U^\prime r - U}{U(\alpha^\prime r - \alpha) + \alpha}  +
(1+v)\frac{\alpha}{r^2-\alpha^2}\right),
\label{eq:ConsistenceTalpha}\\[1mm]
&~& \hspace*{-0.3cm}    \frac{\rho^\prime r}{\rho} = (\alpha^\prime r -
\alpha)
\left( \frac{1 + \frac{U^\prime r}{2} - U}{U(\alpha^\prime r - \alpha)
+ \alpha}
+ \frac{(1+v)}{(r^2-\alpha^2)}\frac{r^2{\dot\alpha}^2\alpha}
{\left( r^2{\dot\alpha}^2-U
({r^2-\alpha^2})\right)}\right).  \label{eq:ConsistenceRalpha}
\end{eqnarray}
%%%%%%%%%%%%%%
In the above, we have also used equation (\ref{ThTh}) to eliminate the quantity
$\kappa_5\rho/6n$ from all other equations. The above system may lead to
either static or time-dependent brane configurations depending on whether 
time-dependence is permitted in the brane trajectory $\chi$. 

\subsection{The Static Brane : An exact solution}

If we assume that the brane trajectory is time-independent, then
considerable simplifications occur. Since $\dot \chi=\dot \alpha=0$, 
${\dot\rho}=0$ from
(\ref{eq:ConsistenceTalpha}). The remaining equations (apart from
(\ref{eq:IsraelTRalpha}) which is trivially satisfied) become:
%%%%%%%%%%%%%%
\begin{eqnarray}
v &=& -(\alpha^\prime r - \alpha) \left( \frac{\frac{1}{2}U^\prime r}
{U(\alpha^\prime r - \alpha) + \alpha} \right),\label{eq:staticTT}\\[1mm]
Ur^2 \alpha^{\prime\prime} &+& (\alpha^\prime r - \alpha) \left(\frac{1}{2}
U^\prime r - U\right) - \alpha = 0,  \label{eq:staticRR}\\[1mm]
U(\alpha^\prime r - \alpha ) + \alpha &=& \frac{\kappa_5}{6}\rho r
\left[U(\alpha^\prime r -
\alpha )^2 +r^2 - \alpha^2\right]^{\frac{1}{2}},
\label{eq:staticthetatheta}\\[2mm]
\frac{\rho^\prime r}{\rho} &=& (\alpha^\prime r - \alpha) \left( \frac{1 +
\frac{1}{2}U^\prime r - U}{U(\alpha^\prime r - \alpha) + \alpha} \right).
\label{eq:staticrestricR}
\end{eqnarray}
%%%%%%%%%%%%%%
Integrating (\ref{eq:staticrestricR}) gives the energy density as:
\begin{eqnarray}
  \rho(r) = \frac{\rho_0}{r^2}\,\left[U(\alpha^\prime r - \alpha) +
\alpha\right],
  \label{eq:staticrho}
\end{eqnarray}
where $\rho_0$ is an integration constant. Substituting for $\rho(r)$ in
(\ref{eq:staticthetatheta}) yields
\begin{eqnarray}
    U(\alpha^\prime r - \alpha )^2 - \alpha^2 + \left(1 - \frac{36}
    {\kappa_5^2\rho_0^2}\right)r^2 = 0\,. \label{eq:staticthetatheta2}
\end{eqnarray}
In fact, equations (\ref{eq:staticRR}) and (\ref{eq:staticthetatheta2}) 
for $\alpha(r)$
can be integrated out in terms of a modified radial variable
\be
{\tilde r} = \int \frac{dr}{r\sqrt{U}} \label{rtilde}
\ee
giving:
\be
\cos\chi = a e^{\tilde r} + b e^{-\tilde r}\,,
\label{gensoln}
\ee
where
\be
4ab=1 - \frac{36}{\kappa_5^2\rho_0^2}\,.
\ee
%with $\tilde r_0$ an arbitrary integration constant following
%from the definition of $\tilde r$ (\ref{rtilde}). 
The energy density and pressure can then be 
straightforwardly obtained as
\bea
\rho = \frac{\rho_0}{r} \left [ \sqrt{U} \left ( 
ae^{\tilde r} - b e^{-\tilde r} \right ) + a e^{\tilde r} + b e^{-\tilde r}
\right ]\,,\label{fullrho}\\
p(r) = - \frac{2}{3} \rho(r) - \frac{\rho_0 U'}{6\sqrt{U}}
\left (ae^{\tilde r} - b e^{-\tilde r} \right )\,.\label{fullp}
\eea
Finally, we can read off the induced metric on the brane as
\be
ds^2 = - U d\tau^2 + \frac{36}{\kappa_5^2\rho_0^2}\,
\frac{r^2 dr^2}{U(r^2-\alpha^2)} + (r^2-\alpha^2)\,d\Omega_{_{\rm II}}^2\,.
\label{induced-family}
\ee
Note that the constants $a$ and $b$ encode the same information as the
integration constant $\rho_0$, and the arbitrary constant corresponding
to the zero point of $\tilde r$ from the integration in (\ref{rtilde}).

%%%%%%%%%%%%%%%%%%%%%%%%%%%%%%%%%%%%%%%%%%%%%%%%%%%%%%%%%%%%%%%%%

\section{Static Braneworld ``Stars''}\label{sec:genstat}

In the previous section, we showed how the static brane equations
admitted an (implicit) exact solution in terms of the radial variable
$\tilde r$, which depended on an integral of the bulk Newtonian potential
$U(r)$. Although this is an exact solution, the actual properties of the
brane depend on the specifics of the relation between $\tilde r$ and $r$.
Once this is determined, we have a solution describing a static, 
spherically symmetric distribution of an isotropic perfect fluid on 
the brane, i.e.\ a solution to the brane TOV
system. Of course, not all the trajectories we will find will
have the interpretation of a star, and, to this end, a careful
examination of the energy density and pressure profile will be
undertaken.

As a warm up to considering more physically relevant bulk spacetimes,
let us consider the simplest possible bulk -- the vacuum: $U=1$. In this
case ${\tilde r} = \ln r$, and $\alpha(r) = a r^2 + b$. 
Introducing the polar coordinates 
\be
x^{*} = r \cos \chi\,, \;\;\;\;
y^{*} = r \sin \chi\,,
\label{polar}
\ee
brings the brane trajectories in the form
\be
\left ( x^* - \frac{1}{2a} \right )^2 + y^{*2} =
\frac{1}{4a^2} - \frac{b}{a}\,,
\ee
with $x^*=b$ in the particular case $a=0$. These solutions
are of limited physical importance as they have constant energy and
pressure\,: $\rho = 2a\rho_0$, $p=-2\rho/3$.
They correspond to an Einstein static universe,
\be
ds^2 = -d\tau^2 + R_0^2 d\Omega_{\rm III}^2
\ee
where $R_0^2 = (1-4ab)/4a^2$.

Now let us consider a more general family of bulk spacetimes. If we
make the choice $U(r)= 1 + Cr^n$ for the bulk metric function, where
$C$ and $n$ are arbitrary constants, we can integrate $\tilde r$
straightforwardly to obtain an analytic solution 
\be
{\tilde r} = \frac{1}{n} \ln \left | \frac{\sqrt{U}-1}{\sqrt{U}+1}
\right|\,,
\ee
which allows us to write $\alpha$ in the general form:
%%%%%%%%%%%%
\begin{equation}
\alpha(r)= \left[A \left|\sqrt{U}-1\right|^{\frac{2}{n}} +
B \left(\sqrt{U}+1\right)^{\frac{2}{n}}\right]\,, \label{sol-static}
\end{equation}
%%%%%%%%%%%%
with $A$ and $B$ convenient redefinitions of the integration
constants $a$ and $b$ appearing in the general solution (\ref{gensoln}).

These solutions (\ref{sol-static}), in conjunction with the choice for the
bulk metric function $U(r)=1+C r^n$, describe different brane
configurations in a variety of spherically symmetric bulk backgrounds.
Two such backgrounds of immediate physical significance correspond
to $n=\pm2$, i.e.\ pure adS spacetime, and the Schwarzschild solution
in five dimensions.

%%%%%%%%%%%%%%%%%%%%%%%%%%%%%%%%%%%%%%%%%%%%%%

\subsection{A 5-dimensional Anti-de Sitter Bulk}

In the case of a 5-dimensional bulk filled with a negative cosmological
constant, the bulk metric function may be written as $U(r) = 1 + k^2r^2$,
where $k$ is the inverse adS radius. The shape of the brane, $\alpha(r)$,
is then given by the expression:
\begin{eqnarray}
\alpha(r) \equiv r\cos\chi(r) = A\left(\sqrt{U}-1\right) +
B\left(\sqrt{U}+1\right)\,, \label{eq:braneADS}
\end{eqnarray}
where, in terms of (\ref{gensoln}), we have set ${\tilde r}=0$ at
infinity, and $A= a/k$, and $B=b/k$.
%%%%%%%%%%%
Using the polar coordinates (\ref{polar}),
the above trajectory may be written as
\be
\left (1-\beta\right)
\left ( x^* + \frac{(A-B)}{1-\beta} \right )^2 - \beta y^{*2} =
\frac{(A+B)^2\,\left(1 - 4ABk^2\right)}{\left(1-\beta \right)}\,,
\label{eq:flatbrane}
\ee
%%%%%%%%%%%%%%%%%%%%%%%%%%
\FIGURE{
\label{fig:ads1}
\includegraphics[height=9.5cm]{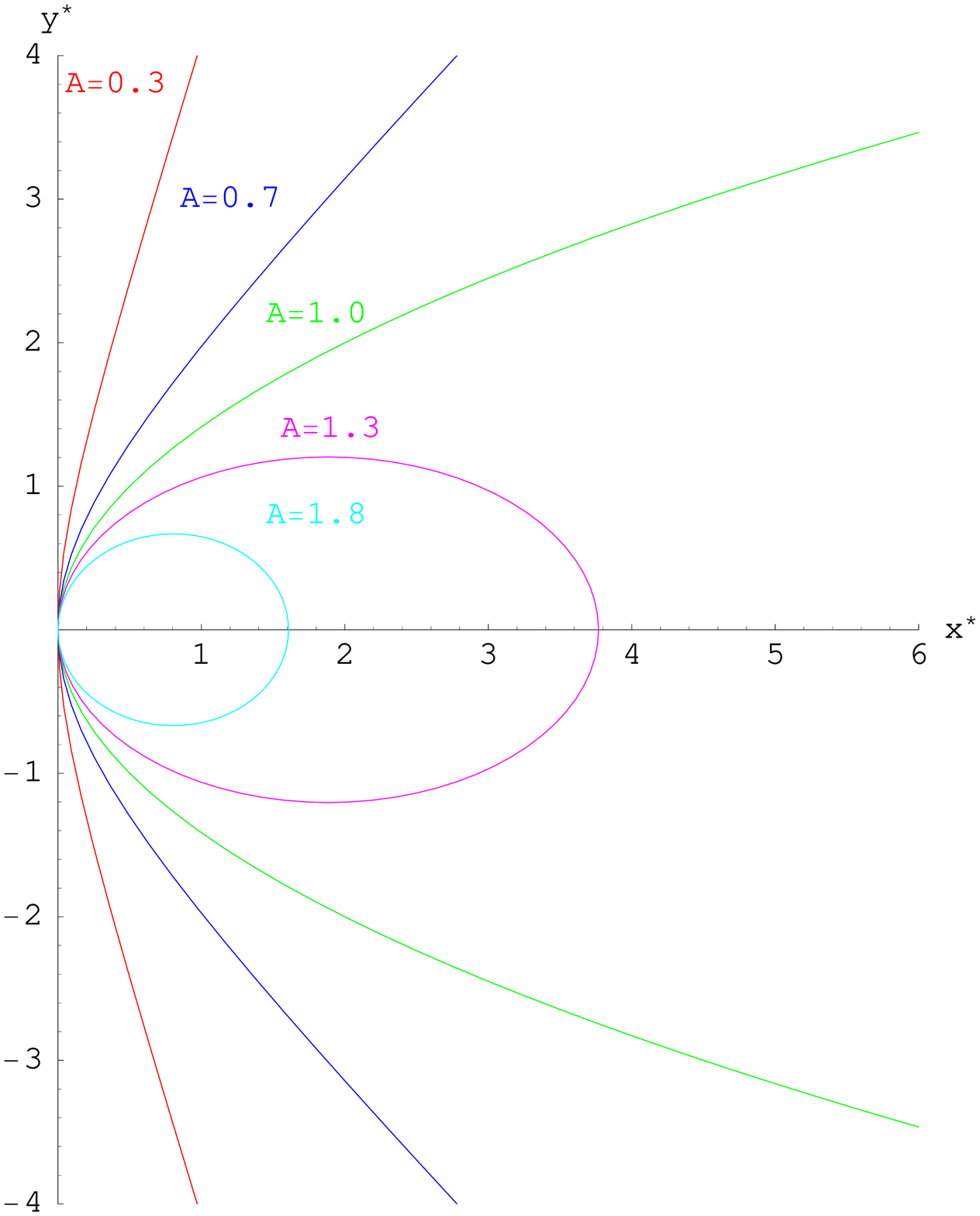}
\caption{A selection of branes of varying coefficient $A$, for the
case $B=0, k=1$ in a 5-dimensional anti-de Sitter bulk. }}
%%%%%%%%%%%%%%%%%%%%%%%%%%
where $\beta=k^2(A+B)^2$. These brane trajectories can be seen to be
conic sections classified by the parameter $\beta$. For $\beta>1$,
the brane is an ellipsoid, $\beta=1$, a paraboloid, and for $\beta<1$
a hyperboloid, with $\beta=0$ corresponding to a straight line.
In figure \ref{fig:ads1}, we display the resulting brane configurations for
some indicative values of the integration parameters $A$ and $B$. For
simplicity, we have set $k=1$, and $B=0$; then, as $A$ varies, the
shape of the brane changes gradually covering
all three cases outlined above.

The physical significance of $\beta$ becomes apparent from the computation
of the energy density from (\ref{eq:staticrho}): 
\be
\rho = k^2\rho_0 (A-B)= \frac{6k}{\kappa_5} \frac{k(A-B)}
{ \sqrt{1 - \beta + k^2 (A-B)^2}}\;.
\ee
This reveals that the energy density is constant throughout the
brane, and, for $A>B$, it remains positive. Then, for the critical value
$\beta=1$, the energy density has precisely the Randall-Sundrum critical
value $\rho_{RS}=6k/\kappa_5$, while for $\beta$ less (greater) than unity
we have a sub- (super-) critical brane. Turning to the equation of
state on the brane, by using equations (\ref{eq:staticTT}) and (\ref{v}), we
find that
%%%%%%%%%%%%
\begin{equation}
p(r)=-\rho + \frac{(A+B) k^2 \rho_0}{3 \sqrt{U}}\,.
\label{p-AdS}
\end{equation}
%%%%%%%%%%%%
It is worth noting that we cannot get a static trajectory for a
critical or super-critical brane with a pure tension energy momentum 
($w=-1$): such a solution follows only for $A+B=0$, that corresponds
to $\beta=0$ and thus to a sub-critical, or Karch-Randall brane \cite{KR}.
For $\beta\neq0$, we have a varying tension on our brane, 
equivalent to a surplus pressure in the braneworld. 

Finally, we can evaluate the induced metric on the brane: 
%%%%%%%%%%%
\bea
ds^2 &=& -U(r)\,d\tau^2 + \left(\frac{1-4ABk^2}
{r^2 - \alpha^2(r)}\right)\frac{r^2dr^2}{U(r)} 
+ \left(r^2 - \alpha^2(r)\right) d\Omega_{\rm _{II}}^2\,,\nonumber \\
&=& - U\left(r({\hat r})\right)\,d\tau^2 + \frac{d{\hat r}^2}
{1-\lambda {\hat r}^2/3} +
{\hat r}^2 d\Omega_{\rm _{II}}^2 \, .
\label{eq:branemetricADS}
\eea
where $\lambda/3=-k^2+\rho^2\kappa^2_5/36$ is the effective cosmological
constant on the brane. Clearly, the spatial part of the metric takes the
form of a constant curvature space, being flat, anti de Sitter or de Sitter
depending on whether the brane is critical, sub- or super-critical
respectively. However, since the relation between $r$ and $\hat r$ is
in general convoluted, the brane has a nontrivial Newtonian potential.
This is because unless $A=-B$, there is a nonvanishing excess pressure
on the brane, which acts as a source and results in a non-asymptotically
flat (or (a)dS) spacetime. To see this explicitly consider a critical 
brane
\be
ds_c^2 = - \frac{k^2\left ( A^2+B^2+{\hat r}^2/2 \right )^2}{(A-B)^2}\,d\tau^2
+ d{\hat r}^2 + {\hat r}^2\,d\Omega_{\rm _{II}}^2\,.
\ee
This spacetime is clearly not asymptotically flat, and in fact corresponds
to a source
\be
T^0_0 = 0\,, \qquad T^r_r=T^\theta_\theta = T^\varphi_\varphi=
\frac{4}{\kappa_5 (2A^2+2B^2+\hat r^2)}\,,
\ee
which corresponds to the actual pressure discrepancy
on the brane: $p+6k/\kappa_5$. 
Similar results hold also for the case of a sub- or super-critical
brane where $T^0_0=\lambda\neq 0$.

We see therefore that these particular trajectories have excess 
pressure on the brane, which results in metrics which do not
asymptote exact Randall-Sundrum or Karch-Randall branes.
However, if $|kA|$ and $|kB|$ are large enough, the metric 
can be flat (or asymptotically (a)dS) over many orders
of magnitude before the effect of the pressure kicks in.

\subsection{A 5-dimensional Schwarzschild Bulk}

We now assume that the 5-dimensional bulk contains not a
cosmological constant, but a mass that
creates a sphe\-rically symmetric Schwarzschild background with
$U(r) = 1-\mu/r^2$, where $\mu$ is related to the actual mass
of the black hole by $M=3\pi\mu/8G_5$. In this case,
$n=-2$ and $C=-\mu$. Then, equation (\ref{sol-static}), that describes
the shape of the brane, takes the form
\begin{eqnarray}
\alpha(r) = r^2\left[A\left(\sqrt{U} - 1\right) +
B\left(\sqrt{U} + 1\right)\right], \label{sol-Schwarz}
\end{eqnarray}
where now $A = -b/\sqrt{\mu}$, $B=a/\sqrt{\mu}$, and ${\tilde r}=0$ at
the horizon.
Note that by construction, these
trajectories are strictly only valid outside the 
event horizon of the black hole, since the definition of 
the ${\tilde r}$ coordinate involves a branch cut there. We could
in principle redefine ${\tilde r}$ inside the horizon, although as 
we are only interested in the exterior solution we shall not do so here.

Following the same analysis as before, the functions $\rho$ and $p$ are now
found to be
\be
\rho(r) = \rho_0 \left [ B(\sqrt{U}+1)^2 
-A(\sqrt{U}-1)^2 \right ] \;, \label{rho-sch}
\ee
\be
p(r) = -\rho(r) + \frac{\rho_0}{3\sqrt{U}} \left [ 
B(\sqrt{U}+1)^2(2\sqrt{U}-1) - A(\sqrt{U}-1)^2 (2\sqrt{U}+1) \right]\;,
\label{p-sch}
\ee
where $\rho_0 = \frac{6}{\kappa_5\sqrt{1+4AB\mu}}$. Clearly $\rho$ is not
a constant for these branes, in contrast to the adS case, and we must
now consider what we think of as a physically sensible brane energy and
trajectory. Obviously we want $\rho$ to be positive, but in addition,
if our solutions are to correspond to stars or black holes, we expect
that they will have the interpretation of energy sources, in other words,
the energy of the brane will increase towards the centre of the brane. 
This is not quite as straightforward as having $\rho$ be a 
decreasing function of $r$, since the brane radial coordinate 
is in fact $r\sin\chi$, and we must therefore examine each trajectory
in turn.

To examine the shape of the brane, we square equation
(\ref{sol-Schwarz}) to obtain:
\be
4AB\,r^2 + 2(B-A)r\cos\chi - \cos\chi^2 = \mu (A+B)^2\,.
\label{eq:braneshapeSCHWARZ}
\ee
The solutions of the above equation are hyperbolae in the $(\cos\chi,r)$-plane,
which leads to the following parametric solution in polar coordinates:
\bea
r &=& \sqrt{\mu} \cosh \lambda \label{rschdef}\\
\chi &=& {\rm Arccos} \sqrt{\mu}(Be^\lambda - A e^{-\lambda})\,.
\label{chischdef} 
\eea
Clearly there are constraints on the range of the parameter
$\lambda$, since we require that $|\cos\chi|\leq1$, and in addition
we will impose positivity of $\rho$.  

With these constraints in mind, we can see qualitatively the 
different families of trajectories that are allowed. First of all, note
that the brane can only touch the event horizon if
\be
0 < B-A \leq 1/\sqrt{\mu}\,,
\ee
the first inequality coming from positivity of energy. Then,
computing the derivative $\frac{d\chi}{dr}$ from (\ref{rschdef},
\ref{chischdef}) shows that unless $B=-A$, $\frac{d\chi}{dr}\to\infty$
at the horizon, meaning that the brane touches the horizon
at a tangent. However, if $B=-A$, then the brane can actually
pass through the horizon, and, as we will see, eventually hits
the central singularity.  

From (\ref{eq:braneshapeSCHWARZ}), we see that the general shape
of the trajectories is in fact primarily determined by the quantity $AB$.
We will now run through the general brane shapes that are
allowed.
%%%%%%%%%%%%%%%%%%%%%%%%%%%%%%%%%%%%%%
\FIGURE{
\includegraphics[height=9.5cm]{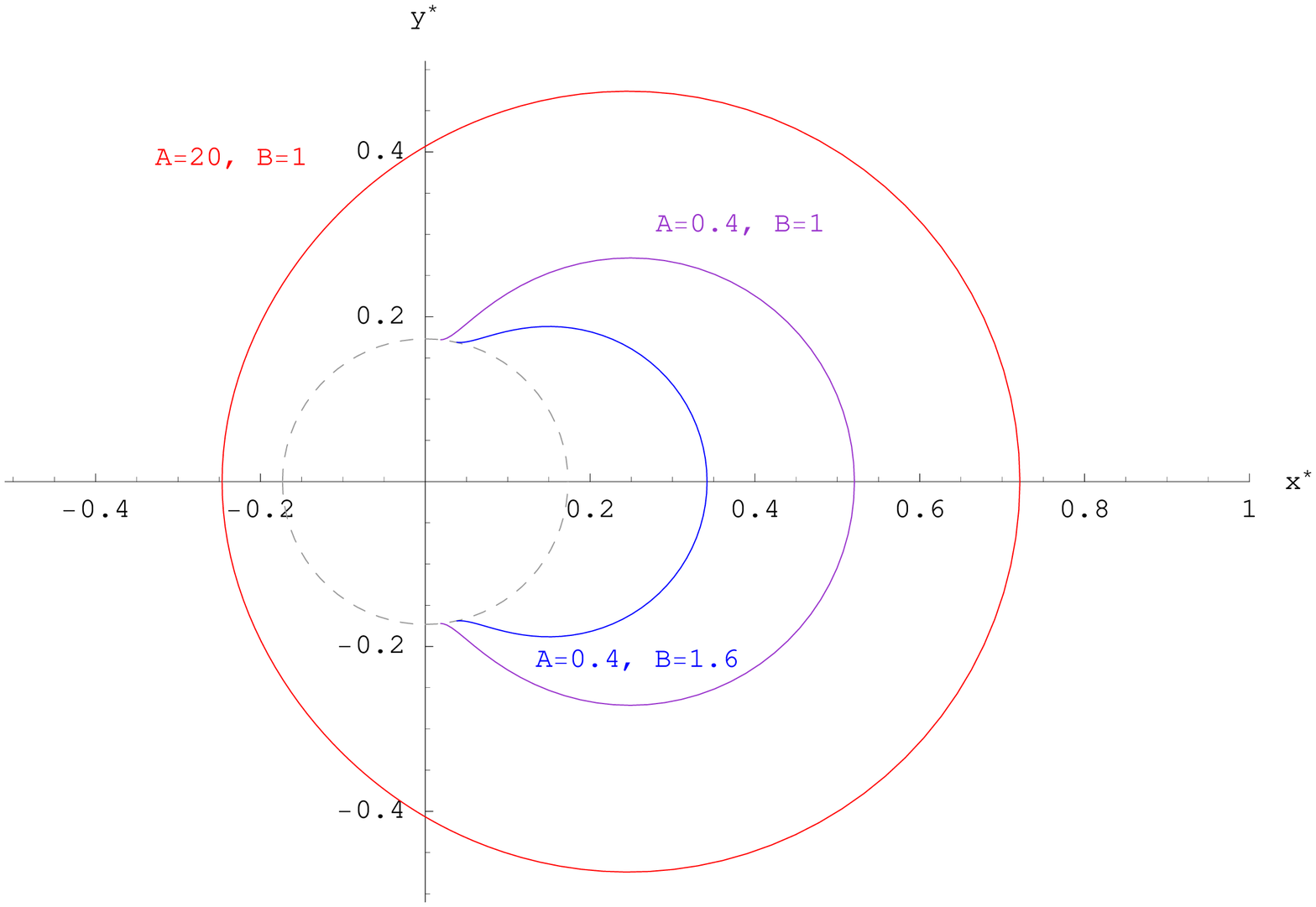}
\caption{A selection of branes (solid lines) for the case $AB>0$,
in a 5-dimensional Schwarzschild bulk of fixed mass parameter
$\mu=0.03$. The dashed line denotes the corresponding horizon radius.}
\label{fig:schwarz2}
}
%%%%%%%%%%%%%%%%%%%%%%%%%%%%%%%%%%%%%%%%%%%

\vskip 2mm

\noindent $\bullet$
$AB>0$. In this case (\ref{chischdef}) implies that
either the brane completely encloses the event horizon, or it
touches (and hence terminates on) the event horizon, depending on
the value of $B-A$ compared to the critical value $1/\sqrt{\mu}$.
Returning to 
(\ref{sol-Schwarz}), we see that in the latter case $\cos\chi$ is an 
increasing function of $r$, and so the brane lies to the right
of the points where it touches the event horizon, which are themselves
on the right of the $(x^*,y^*)$-plane since $B>A$. Figure \ref{fig:schwarz2}
shows a sample set of brane trajectories in this class.
Note that as $\chi'<0$ for these trajectories, it is
the interior of the bubbles in the bulk that is retained.

For $B>A>0$, it may easily be seen that the energy density (\ref{rho-sch})
remains positive throughout the brane. Turning to (\ref{eq:staticrestricR}),
we see that this also corresponds to $\rho$ being an increasing function of
$r$, hence these branes have an energy surplus at the point located farthest
away from the event horizon. The branes terminating on the horizon look
like the inside of a bubble, with the event horizon defining its boundary,
and its energy density concentrated at its center $\hat r=0$. 
However, the pressure (\ref{p-sch}) is increasing away from the center of
the bubble, eventually acquiring an infinite value at the point where the
brane touches the event horizon thus rendering the bubble boundary singular. 

%%%%%%%%%%%%%%%%%%%%%%%%%%%%%%%%%%%%%%
\FIGURE{\hspace*{1cm}
\includegraphics[height=10cm]{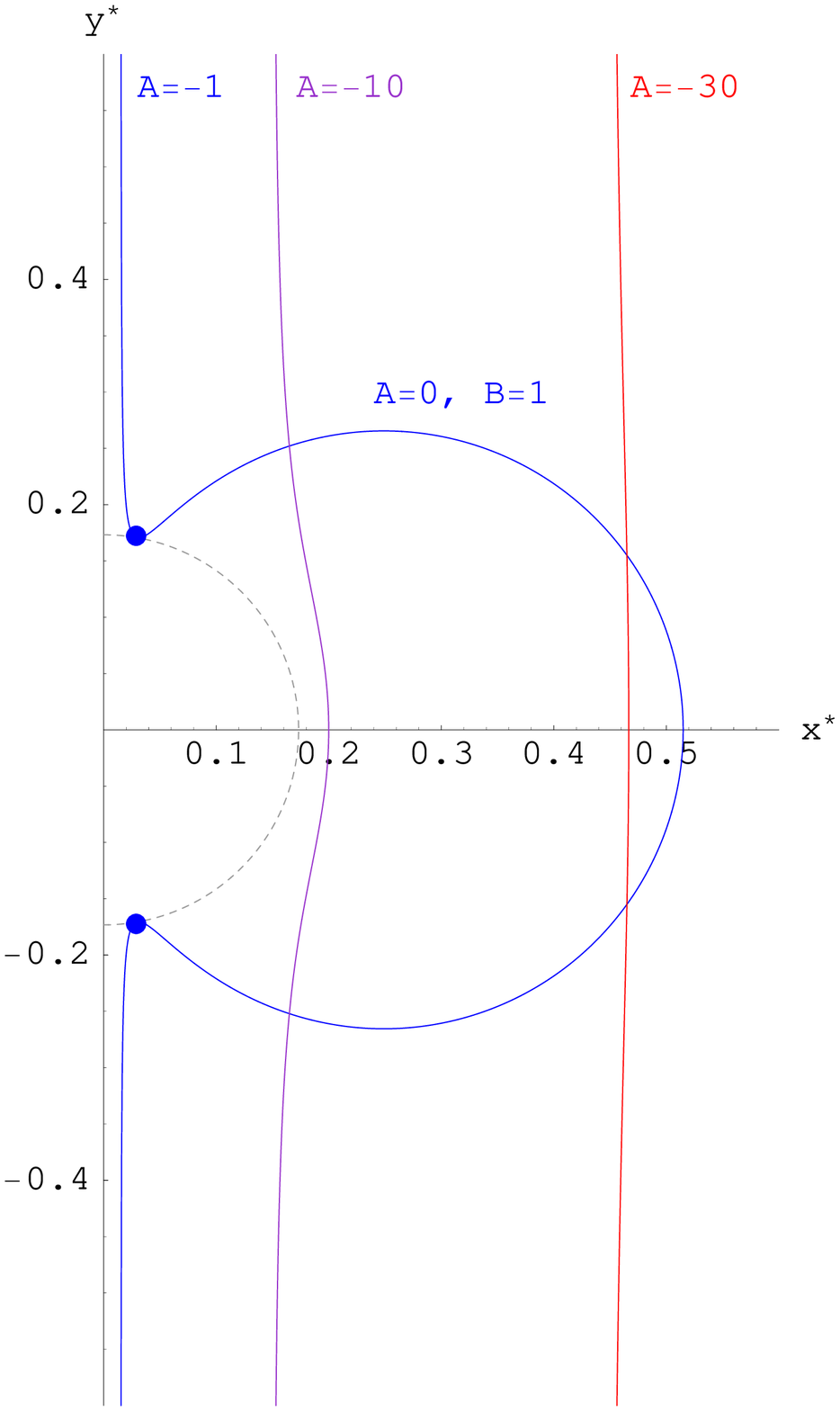}
\caption{A selection of branes for the case $AB=0$, in a 5-dimensional
Schwarzschild bulk of fixed mass parameter $\mu=0.03$. The case $A=0, B=1$
is shown together with a set of branes with $B=0$ and variable $A$. 
The dashed line denotes again the event horizon. 
}
\label{fig:schwarz1}
}
%%%%%%%%%%%%%%%%%%%%%%%%%%%%%%%%%%%%%%%%%%%

\vskip 2mm
%\item 
\noindent $\bullet$
$AB=0$. If $A=0$, then (\ref{sol-Schwarz}) will only correspond to a
brane exterior to the horizon if $|\cos\chi|=rB\,(1+\sqrt{U}) \leq 1$.
This leads to the bound $B \leq 1/\sqrt{\mu}$ while positivity of
energy demands that $B>0$.  Examination of (\ref{sol-Schwarz})
then shows that these trajectories start tangent to the event
horizon, curve out into the bulk, then return to the event horizon.
The indicative case ($A=0$, $B=1$) is shown in figure \ref{fig:schwarz1}.
The energy density and pressure of these trajectories may be
easily shown to be similar to the ones of branes with $AB>0$
terminating on the horizon, and they correspond again to bubbles
with a singular pressure boundary. 

If $B=0$ on the other hand, the trajectories asymptote $r\cos\chi=
-\mu A/2$ at infinity, remaining roughly straight until they near the
vicinity of the horizon where they bend away. (Those touching the
horizon are the planar solutions of \cite{Seahra}.)
If $-A<1/\sqrt{\mu}$, then they become tangent to the horizon 
at the same point as the trajectory, found by swopping $B$ and $-A$.
If $-A>1/\sqrt{\mu}$, the trajectories
manage to bend sufficiently far that they avoid the event horizon
altogether. The borderline case $A=-1/\sqrt{\mu}$ has the brane just
skimming the horizon. Figure \ref{fig:schwarz1} shows some of
these brane trajectories with $B=0$ and variable $A$.
%%%%%%%%%%%%%%%%%%%%%%%%%%%%%%%%%%%%%%
\FIGURE{
\hspace*{1cm}\includegraphics[height=9.5cm]{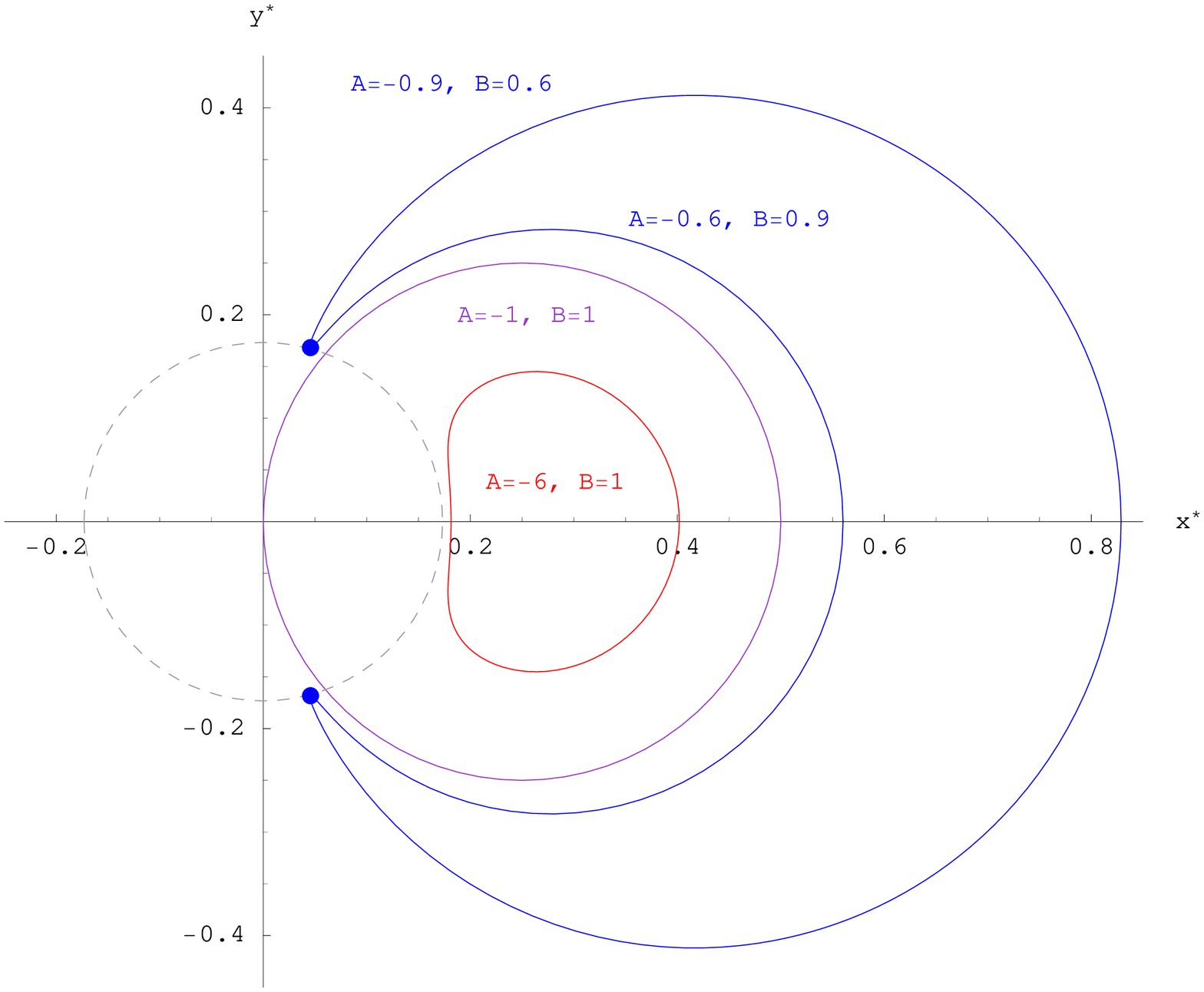}
\caption{A selection of branes for the case $AB<0$,
in a 5-dimensional Schwarzschild bulk of fixed mass parameter
$\mu=0.03$. The dashed line denotes again the corresponding
horizon radius.}
\label{fig:schwarz3}
}
%%%%%%%%%%%%%%%%%%%%%%%%%%%%%%%%%%%%%%%%%%%

The energy and pressure of these brane trajectories have a particularly
simple form\,:
\be \label{schstarem}
\rho = \frac{-6A}{\kappa_5} \left ( \sqrt{U}-1 \right )^2\,,
\qquad p = - \frac{\rho}{3} \frac{(\sqrt{U}-1)}{\sqrt{U}}\,.
\ee
For $A<0$, these branes have $\rho$ positive and uniformly decreasing
as $r$ increases.
If $|A|\leq 1/\sqrt{\mu}$, the energy density decreases away
from the horizon, however the pressure still diverges there. If
$|A|> 1/\sqrt{\mu}$, the brane never touches the horizon and the
pressure remains everywhere finite. Moreover, the energy density is
decreasing away from their centre $\hat r=0$, and hence they correspond
to asymptotically empty branes with positive mass sources. We will
return to these physically interesting cases later.

\vskip 2mm
\noindent $\bullet$
$AB<0$. From positivity of energy, we see that here $A$ must be
negative, and the brane lies exclusively on the right hand side
of the $(x^*,y^*)$-plane. The brane can be seen to either be a
single arc which touches the horizon, or a closed loop. In fact,
by slowly reducing the magnitude of $B-A$, we see that the loop
moves closer to the event horizon forming two arcs, each curving
away from the horizon, eventually touching it at two points. In
this case, the outer half is related to the inner half by exchanging
$B\to -A$, and $A\to-B$. A selection of these branes is depicted
in figure \ref{fig:schwarz3}. 

For the arc branes with $B-|A|>0$,
the energy density increases away from the horizon and reaches its
maximum value at the center of the brane $\hat r=0$, creating again
a bubble with a diverging pressure at its boundary. The same 
singular behaviour at the event horizon is exhibited by the
pressure in the case of arc branes with $B-|A|<0$, although in
this case, the energy initially decreases away from the horizon,
which corresponds to an energy deficit, and then increases again
creating an energy source. In the case of closed loops, that do
not touch the horizon, the pressure singularity is again avoided, and
the energy density profile resembles one of the two described above
depending on the value of $B-|A|$ in that case.

The one exception to the picture described above, where the brane can
at most touch the horizon but otherwise it extends outside it, is the
special situation alluded to previously, namely the case $A=-B$. 
The brane equation in this case can be written as
\be
{y^{*}}^2 + \left(x^{*}-\frac{1}{4B}\right)^2 = \frac{1}{16B^2}\,.
\ee
The above situation is unique in that the brane extends beyond the black
hole horizon and even passes through the point mass located at
$x^*=y^*=0$. For this solution the singularity problem associated with
crossing the horizon is removed by the choice $A=-B$. The energy
density is uniformly increasing with $r$ for $B>0$, and acquires its
maximum value at the point located farthest away from the black hole.
This type of brane is also shown in figure \ref{fig:schwarz3} --
the solid purple line indicating the sole trajectory which crosses
the horizon.

Having derived these brane trajectories, we would now like to highlight
which are likely to be physically useful. First of all, note that the
definition of the brane extrinsic curvature means that the normal,
defined in (\ref{basis}), is pointing out of the spacetime being
kept in this $\mathbb{Z}_2$-symmetric identification. This means that,
typically, for a trajectory which escapes to infinity, it is the 
{\it right} hand side of the bulk spacetime which is being kept, 
and for closed branes, it is the {\it interior} of the bubble. In
other words, for the brane trajectories with $B=0$, the spacetime
without the black hole in is the bulk appropriate to the brane
trajectory. Similarly, with the small (red) 
bubble in figure \ref{fig:schwarz3},
it is the interior of the bubble which is kept, which has no segment
of the event horizon in it.

Focussing on the $B=0$ trajectories, as these are asymptotically
flat, we now show that these have precisely the energy-momentum one
would expect for a TOV star solution. 
From (\ref{schstarem}) we see that the energy density is 
peaked around ${\hat r}=r\sin\chi=0$, as is the pressure, the energy
falling off as $1/r^4$ and the pressure as $1/r^6$ (with ${\hat r}\propto r$
for large $r$). Plotting 
the energy and pressure for the brane shows that this does indeed
correspond to a localized matter source, with the peak energy density
dependent on the minimal distance from the horizon. 
The central energy and pressure can be readily calculated from this
minimal radius, $r_m = \mu |A|/2 + 1/2|A|$:
\be
\rho_c = \frac{24|A|}{\kappa_5(1+\mu A^2)^2}\,, \qquad
p_c = \frac{16|A|}{\kappa_5(\mu A^2-1)(1+\mu A^2)^2}\,,
\ee
which shows that the central pressure diverges as $\mu A^2\to1$. However, 
for $|A|=1/\sqrt{\mu}$ the trajectory just touches 
the event horizon of the black 
hole, which is the source of this divergent pressure. This is analogous
to the divergence of central pressure in the four-dimensional TOV
system, which is indicative of the existence of a Chandrasekhar limit
for the mass of the star. 
%%%%%%%%%%%%%%%%%%%%%%%%%%%%%%%%%%%%%%
\FIGURE{
\includegraphics[width=5cm]{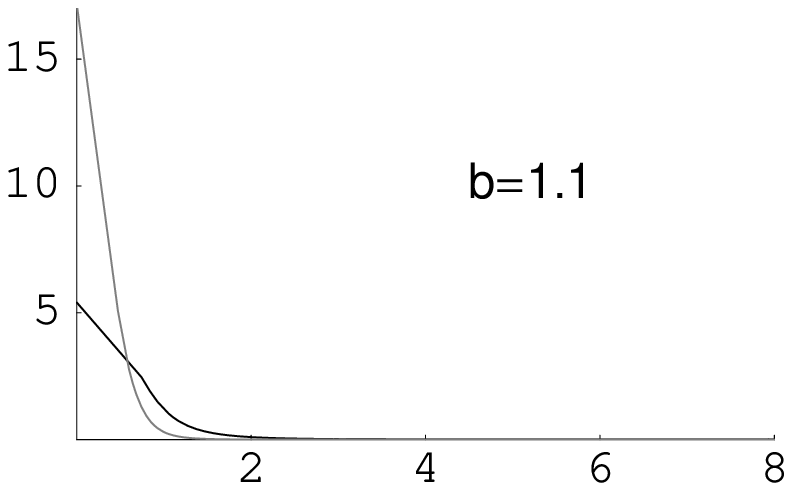}\nobreak\hskip3mm\nobreak
\includegraphics[width=5cm]{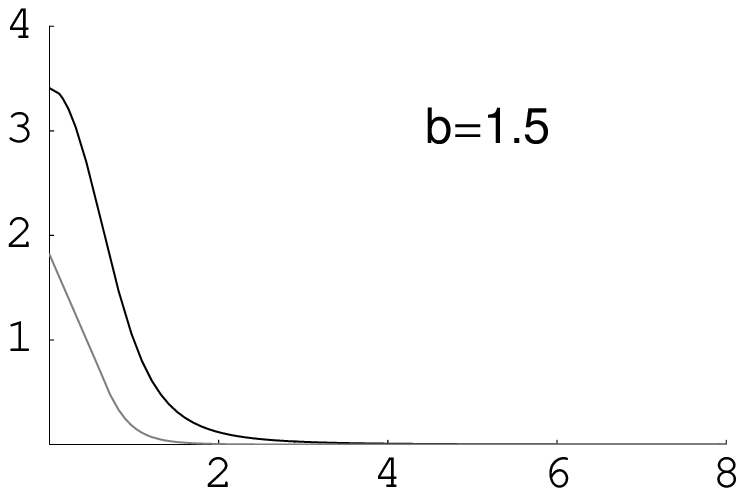}\nobreak\hskip3mm\nobreak
\includegraphics[width=5cm]{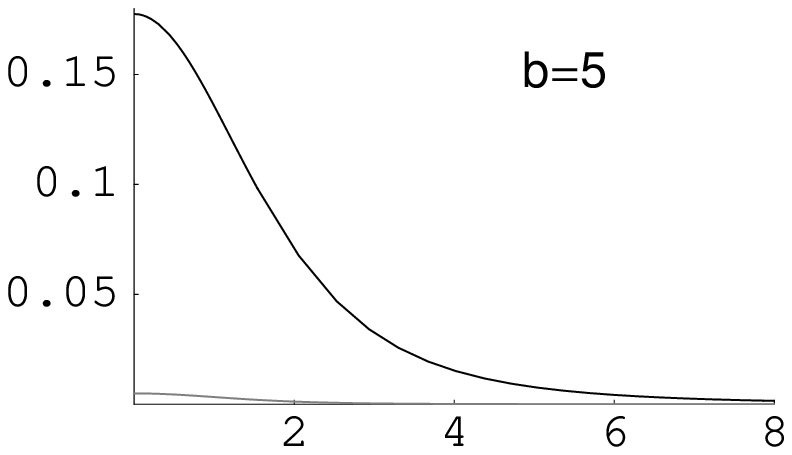}
\caption{The energy (dark line) and pressure (grey line) of 
brane stars with a pure Schwarzschild bulk as a function of the
brane radial coordinate ${\hat r}$. The black hole mass
is fixed at $\mu=1$, and the distance of closest approach to
the horizon increases across the plots.}
\label{fig:schTOV}
}
%%%%%%%%%%%%%%%%%%%%%%%%%%%%%%%%%%%%%%%%%%%

Some examples of the solutions to the brane TOV equations are 
given in figure \ref{fig:schTOV}.  Notice that as the brane
trajectory is moved away from the horizon, the pressure of the matter
on the brane decreases; in fact, in the final plot, the pressure can
barely be distinguished from the axis. Also note that as $\mu A^2$
increases, the spread of the matter on the brane increases.
In these spacetimes, there is no actual black hole in the bulk, since
it is the bulk to the right of the brane that is retained. Rather,
it is the combination of the bulk Weyl curvature and the brane bending
which produces the fully coupled gravitational solution.

By working in dimensionless units, $\xi = r/\sqrt{\mu}$ and $b$, 
we can be more explicit about the dependence of energy on the
bulk Weyl mass parameter, $\mu$, and the `impact parameter' $A$
(or $b$). For instance, the energy density on 
the brane scales as $b/\sqrt{\mu}$,
and the spread of the star as $b$. We can also calculate the total 
mass of the star readily as
\be
M_s = 4\pi 
\int_{r_m}^\infty r\rho\,\frac{\sqrt{r^2-\alpha^2}}{\sqrt{U}}\,dr
= \frac{12\pi\mu}{\kappa_5}\,F(b)\,,
\ee
where $F(b) \to 1$ from above very rapidly. In other words, 
the total mass of the
star is directly proportional to the bulk mass parameter.
Referring back to the existence of a central singularity in the 
pressure, we see that this corresponds to either reducing $b$ to bring
the brane to the horizon, or increasing $\mu$, to bring the horizon
to the brane (or perhaps a combination of both). In either case, the 
total mass and the concentration of the matter increases. There is
however no overall upper bound on the mass of the star, as we can always
have a nonsingular solution for any $\mu$ simply by making $b$ large
enough. The limit on mass is therefore not a true Chandrasekhar limit, but
more a statement about an upper bound on the concentration of matter.
The real reason there is no absolute upper bound is because, unlike the RS
system with an adS bulk, gravity on the braneworld is not localized, nor is
it four-dimensional.
This is also seen in the induced metric on the brane (\ref{induced-family}),
which in the case of a Schwarzschild bulk has no convenient expression in
terms of the radial coordinate ${\hat r}^2 = r^2 - \alpha^2$.
However, for the one solution which tends to infinity, we see that
$\alpha \to \mu |A|/2$, which implies that the asymptotic metric is in fact
the projection of the 5-dimensional Schwarzschild metric on the brane.

To sum up: the pure Schwarzschild spacetime has a rich set of brane
trajectories, most of which are closed, however, there is a class
of asymptotically flat branes which have a localized source
satisfying the DEC (mostly) and hence with the interpretation of an
isolated gravitating star.

%%%%%%%%%%%%%%%%%%%%%%%%%%%%%%%%%%%%%%%

\section{Braneworld Stars : A Schwarzschild-adS Bulk}\label{sec:star}

We now turn to the case of a static brane embedded in a
5-dimensional Schwarzschild--anti de Sitter (Sch-adS) spacetime. Since the
Randall Sundrum model is a brane in adS spacetime, we expect that
any consistent brane trajectories in Sch-adS will potentially correspond
to brane stars or black holes. It is worth stressing that these solutions
will not just be brane solutions, but full brane {\it and bulk} solutions,
since the full Israel equations for the brane have been solved in 
a known bulk background. 

Note that in 
the Randall-Sundrum scenario, our empty brane solution 
does not have zero energy-momentum, but a background constant 
energy and tension: $T_{\mu\nu} = \rho h_{\mu\nu}$, where
$\rho = \rho_{RS}$ for the critical RS brane, and $\rho<\rho_{RS}$ for
the subcritical Karch-Randall brane. 
Therefore, when we compute $p$ and $\rho$ for
a spherically symmetric brane, this background brane energy momentum
will be included. According to \cite{SMS},
the brane gravitational field couples to the {\it differential}
energy momentum
\be
{\cal T}_{\mu\nu} = T_{\mu\nu} - \frac{\rho_b}{\kappa_5}h_{\mu\nu}
\ee
where $\rho_b$ is the background brane tension, as well as a possible
brane cosmological constant term corresponding to the difference between
$\rho_b$ and the critical RS brane tension.
For the critical brane, ${\cal T}_{\mu\nu}$ will give the extra matter
on the brane which is sourcing the spherically symmetric gravitational
field, however, sub- and super-critical branes will also have an
additional gravitational effect from the cosmological constant term. 
For the sake of simplicity, as well as correspondence with the 
zero bulk black hole mass limit, 
we will identify the background brane tension as
\be
\rho_b = \frac{6k(a-b)}{\kappa_5\sqrt{1-4ab}}
\ee

The metric function $U(r)$ is now given by
\begin{eqnarray}
U(r) = 1+k^2r^2 - \frac{\mu }{r^2}\,, \label{metric-SAdS}
\end{eqnarray}
and is not covered by the general metric ansatz studied in the previous
section.  The function $\tilde r$ actually has an exact analytic expression
%%%%%%%%%%%%%%
\begin{equation}
{\tilde r} (r) = \frac{1}{k r_+}\, 
{\rm Elliptic\,F}\left[{\rm Arcsin}\left(\frac{r}{r_-}\right),
\frac{r_-^2}{r_+^2}\right]\,.
\label{exact}
\end{equation}
%%%%%%%%%%%%%%
In the above, $r_+$ (the black hole horizon) and $r_-$ are defined by
the expressions
%%%%%%%%%%%%%
\begin{equation}
r_+^2=\frac{-1+\sqrt{1+4k^2 \mu }}{2k^2}\,, \qquad
r_-^2=\frac{-1-\sqrt{1+4k^2 \mu }}{2k^2}\,.
\end{equation}
%%%%%%%%%%%%%

Although an exact solution, the expression (\ref{exact}) is of limited
use because of the imaginary value of $r_-$, and the presence of the
Elliptic function. It can be used of course to give numerical solutions
for the brane,  as we will do presently, however we will first deduce
some general properties of trajectories based in part on what we have
learned from the effects of the black hole mass, and negative cosmological
constant separately.

First of all, note that for large enough $r$, the geometry will be
dominated by the cosmological constant, therefore we expect that our
pure adS solutions will be good approximations to any trajectories for
large $r$. Next, if $\mu k^2 \ll 1$, i.e.\ if
the black hole is much smaller than the adS scale,
we expect that in the vicinity of the horizon the Schwarzschild 
solutions will be good approximations for the brane, therefore for
small mass black holes, we might expect brane trajectories to be
well approximated by some combination of Schwarzschild and adS
branes. We would like to note here that, for convenience and easy
comparison with the pure adS limit, we
zero the $\tilde r$-coordinate at infinity.
Then, the range of $\tilde r$ in Sch-adS turns out to be finite, and 
to decrease sharply with increasing $\mu$ (for example, if
$\mu=0.01$, $r_+ \simeq 0.1$ and ${\tilde r}_+ \simeq -3.7$, whereas if 
$\mu = 10^5$, $r_+ \simeq10$ and ${\tilde r}_+ \simeq -0.13$).
This suggests that trajectories in large mass Sch-adS black hole
spacetimes are more finely tuned, and possibly more restricted than
in small mass black hole spacetimes.

%%%%%%%%%%%%%%%%%%%%%%%%%%%%%%%%%%%%%%
\FIGURE{
\includegraphics[width=10cm]{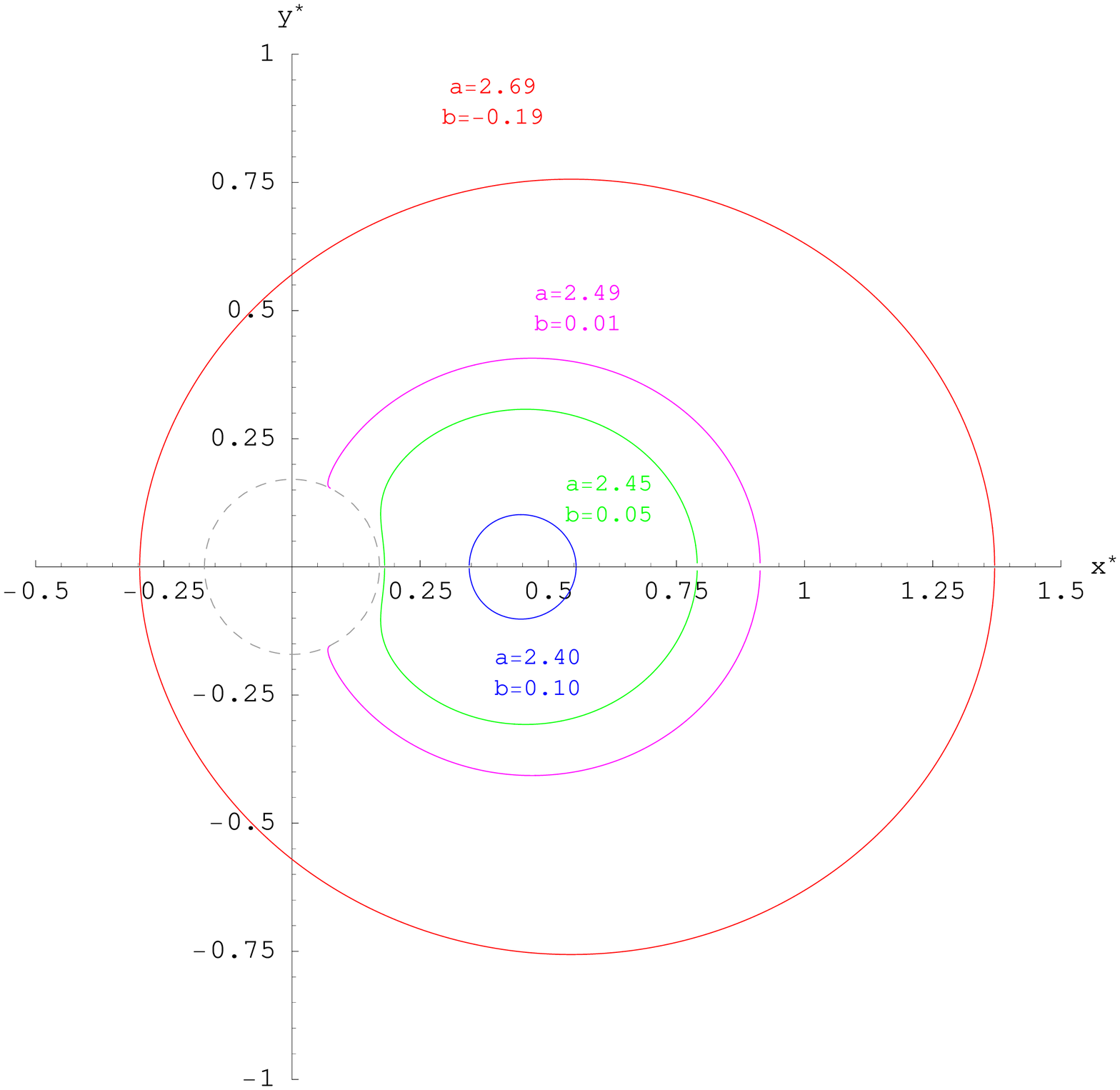}
\caption{A sample of supercritical brane trajectories with $a+b>1$ in a 
5-dimensional Schwarzschild--anti de Sitter background of fixed
parameters $k=1$ and $\mu=0.03$. The dashed line denotes again the
horizon.}
\label{fig:supercrit}
}
%%%%%%%%%%%%%%%%%%%%%%%%%%%%%%%%%%%%%%

We are clearly interested in branes which have matter that can be
interpreted as a gravitating source, i.e.\ we would like to have a
energy excess at the center of the brane. We can see therefore
that, unless we have a closed bubble, this will in general correspond
to $\rho$ being an increasing function of $r$. From (\ref{eq:staticrestricR}),
we see that in Sch-adS
\be
\rho' = \frac{2\mu\rho_0}{r^3}\,(\cos\chi)'\,,
\ee
hence $\rho$ will be a decreasing function of $r$ if $\cos\chi$ is.
However, from (\ref{fullrho}), $\rho$ is asymptotically dominated by 
$\rho_0\sqrt{U}(ae^{\tilde r} - be^{-{\tilde r}})/r \propto (\cos\chi)'$,
hence any positive energy brane trajectory will have $(\cos\chi)'>0$
near infinity, and hence $\rho$ will be increasing near infinity (albeit
at a very slow rate), 
corresponding to an energy deficit at large $r$. However,
this underdensity will prove to be extremely marginal, and many
trajectories have, as their main feature, energies significantly
in excess of their background value in the interior.

Like adS spacetime, the Sch-adS trajectories can be classified according
to whether they asymptote the adS boundary at nonzero $\chi$, at
$\chi=0$, or do not reach the boundary at all, i.e.\ are closed 
bubbles. These correspond to subcritical, critical,
or supercritical branes ($a+b>1$, $a+b=1$, and $a+b<1$) respectively.

\subsection{Supercritical branes}

We will now show that all closed trajectories are supercritical. Clearly
if a trajectory is not closed, it is not supercritical, since if
a brane asymptotes the adS boundary, then in that asymptotic
regime it must have $|\cos \chi| \simeq |a+b| \leq 1$ from the pure
adS results.  Now suppose that the brane has a finite extent.
In that case, the brane must satisfy $\cos\chi=1$ at some value
${\tilde r}_1 < 0$. If the brane is closed, it should also satisfy
$\cos\chi=\cos\chi_0$ at some other value ${\tilde r}_0< \tilde r_1$, 
where $\chi_0=0$ if the bubble lies entirely on the RHS of the black hole,
and $\chi_0=\pi$ if the bubble encloses the black hole. The only
other possibility for a finite brane is to terminate on
the horizon and, in that case, ${\tilde r}_0={\tilde r}_h$ 
if $\chi_0 \in (0,\pi)$.
Using this information, we find that
\be
a = \frac{e^{{\tilde r}_1} - e^{{\tilde r}_0}\cos\chi_0}
{e^{2{\tilde r}_1} - e^{2{\tilde r}_0}}\,, \qquad
b = e^{{\tilde r}_1+{\tilde r}_0}
\frac{(e^{{\tilde r}_1}\cos\chi_0 - e^{{\tilde r}_0})}
{e^{2{\tilde r}_1} - e^{2{\tilde r}_0}}\,.
\ee
From the above we may see that, since ${\tilde r}_0<{\tilde r}_1<0$, 
$a+b$ is a decreasing function of ${\tilde r}_1$; therefore, since $a+b=1$
for ${\tilde r}_1=0$, $a+b>1$ for a closed bubble with $\tilde r_1<0$.

In figure \ref{fig:supercrit}, we depict a sample of supercritical
branes for fixed background parameters $k=1$ and $\mu=0.03$, and various
values of the parameters $a$ and $b$ of the general solution (\ref{gensoln}).
The branes exhibit the features discussed above and form
either closed loops or arcs terminating on the horizon. The latter
characteristic is determined by whether $a$ and $b$ satisfy the
constraint $|\cos \chi| \simeq |a e^{\tilde r_+} + b e^{-\tilde r_+}| \leq 1$
near the horizon. For the arcs terminating on the horizon, it may be
easily seen that the energy density remains positive and increases
towards the center of the brane. As in the Schwarzschild case, however,
the pressure becomes singular at the horizon. For branes forming a closed
loop on the RHS of the horizon, a similar behaviour is found to the
one encountered in the Schwarzschild background: the energy density
decreases near the vicinity of the horizon but increases towards the
most distant point of the brane. A uniformly increasing behaviour for
the energy density is found also in the case of brane trajectories
that enclose the black hole horizon: $\rho$ reaches its maximum positive
value at the point of the brane located farthest away from the black hole,
although care must be taken over the choice of $a$ and $b$ to ensure
that $\rho$ remains positive throughout the trajectory.

%%%%%%%%%%%%%%%%%%%%%%%%%%%%%%%%%%%%%%
\FIGURE{
\includegraphics[height=8cm]{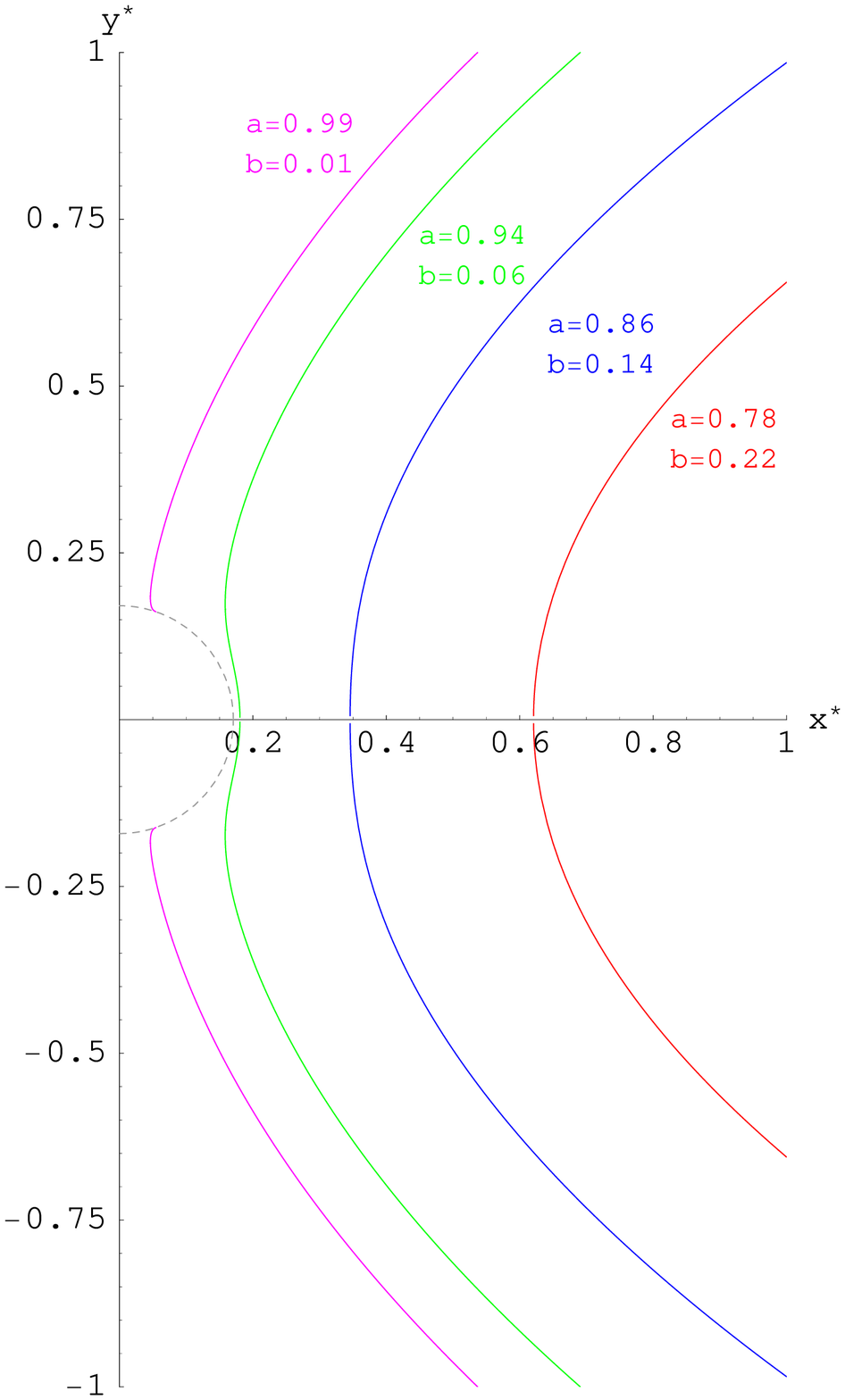}\nobreak\hskip3mm\nobreak
\includegraphics[height=85mm]{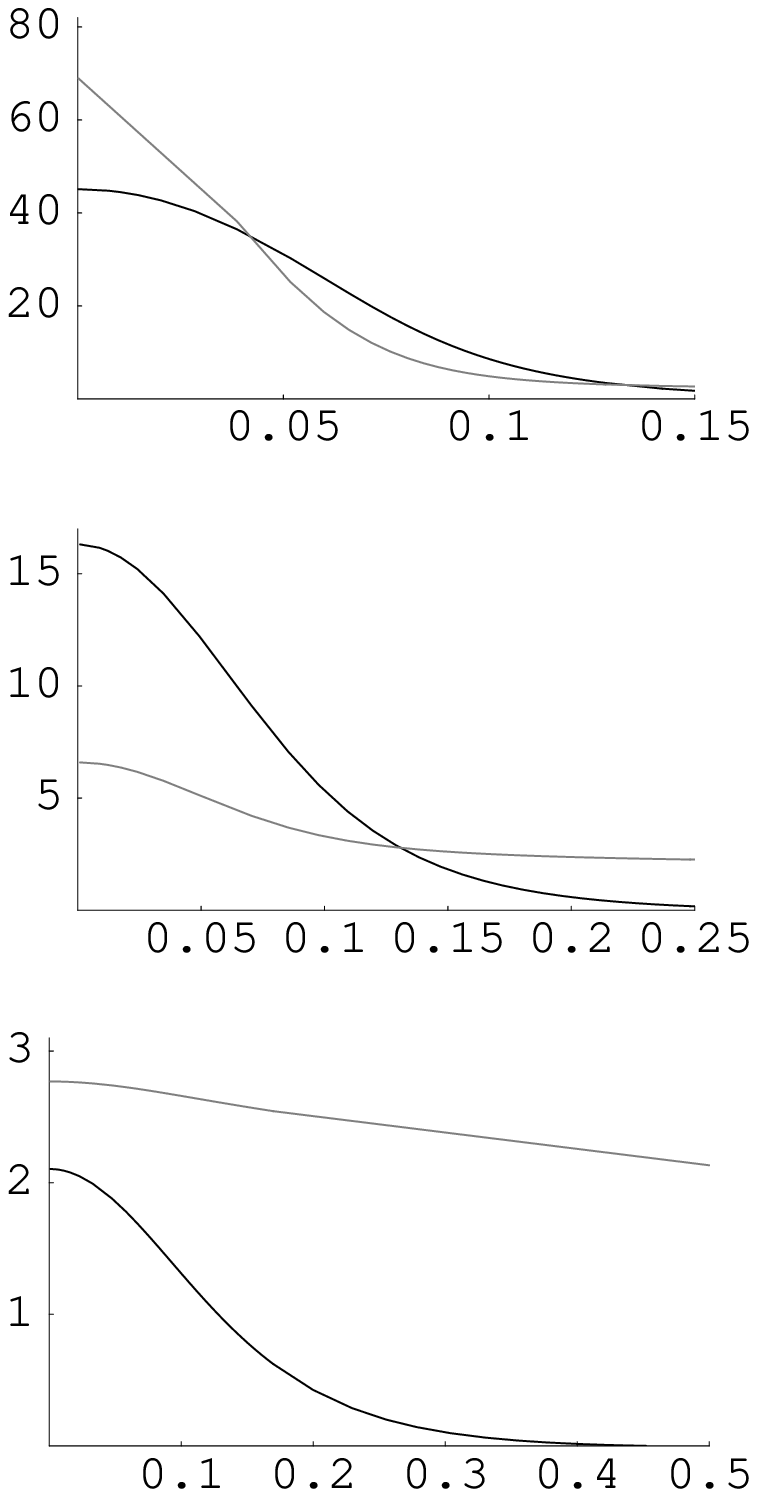}
\caption{(a) A sample of critical brane trajectories with $a+b=1$ in a 
5-dimensional Schwarzschild--anti de Sitter background of fixed
parameters $k=1$ and $\mu=0.03$. The dashed line denotes again the
horizon. (b) A set of plots of the brane energy (black line) and
pressure (grey line) for a sequence of critical branes moving 
away from the horizon.}
\label{fig:crit}
}
%%%%%%%%%%%%%%%%%%%%%%%%%%%%%%%%%%%%%%

\subsection{Critical branes}

In this case we have $a+b=1$, which means that the brane trajectories
asymptote the adS boundary at exactly $\chi=0$. The branes are thus
open, and may or may not touch the black hole horizon depending on
the exact values of the parameters $a$ and $b$. In order to demonstrate
when this happens, we describe our trajectories in terms of a
sole parameter by writing $a=(1+c)/2$ and $b=(1-c)/2$. Then, if
\be
c < |\tanh{\tilde r}_+/2|\,,
\ee
the trajectory will remain on the RHS of the horizon: after reaching
a point of closest proximity, the brane will bend to avoid the horizon and
eventually escape to infinity. If $c$ saturates or exceeds the
above bound, the brane will terminate on the horizon. A sample of
critical trajectories in a Sch-adS background is shown in figure
\ref{fig:crit}(a). 

The behaviour of the energy density and pressure in this case is
strongly dependent on the location of the brane compared to the
black hole. Branes that originate from the horizon and extend to
infinity have their energy density positive, provided they intersect
the horizon at $\chi_0<\pi/2$, i.e.\ $c<|\coth{\tilde r}_+|$. The
horizon is then a local energy maximum if $c < |\tanh{\tilde r}_+|$,
with the energy decreasing away from the horizon, undershooting the
asymptotic value before increasing again towards the asymptotic
critical value. If $c > |\tanh{\tilde r}_+|$, then the energy 
monotonically increases out to infinity.
In addition, the pressure, as expected, diverges at the black hole boundary. 

According to figure \ref{fig:crit}(a), as the value of the $c$ parameter
decreases, the brane shifts towards the right. For branes that 
avoid the horizon the energy density is again positive, peaking
at the center, and dropping rapidly to the background value, 
undershooting it slightly to form the underdense region already discussed.
The pressure also reaches its maximum value at the center, but
is uniformly decreasing with $r$, at a much slower rate, consistent with
the pressure excess observed for the pure adS branes. Apart from this
pressure excess, the other main difference with pure Schwarzschild 
trajectories, is whether branes satisfy the differential DEC at
their center depends crucially on the choice of $c$. By differential
DEC, we mean the DEC for the differential energy momentum ${\cal T}_{\mu\nu}$,
and thus the energy momentum tensor for an observer on the brane. In pure
Schwarzschild, the DEC is satisfied except for branes which skirt extremely
close to the horizon, where the local Weyl curvature causes the pressure
to diverge. This phenomenon is also observed for the Sch-adS branes skimming
close to the horizon, however, as we decrease $c$ (or increase $b$) 
the central energy dominates the pressure for only a finite range of
$b$ before once again dropping below the pressure. This is because the 
further we move away from the horizon the adS curvature becomes more
important, and for pure adS branes, the effect of the adS curvature is 
to induce a pressure excess.
In figure \ref{fig:crit}(b), we present the (differential) energy 
density and pressure for a sequence of critical branes in a 
Sch-adS background displaced an increasing distance from the horizon. 
One may easily observe the localization
of both the energy density and pressure at the center of the brane,
corresponding to a distribution of a positive mass source, as well
as the fact that the DEC can be satisfied at the centre of the
distribution.

%%%%%%%%%%%%%%%%%%%%%%%%%%%%%%%%%%%%%%
\FIGURE{\hspace*{1cm}
\includegraphics[width=8cm]{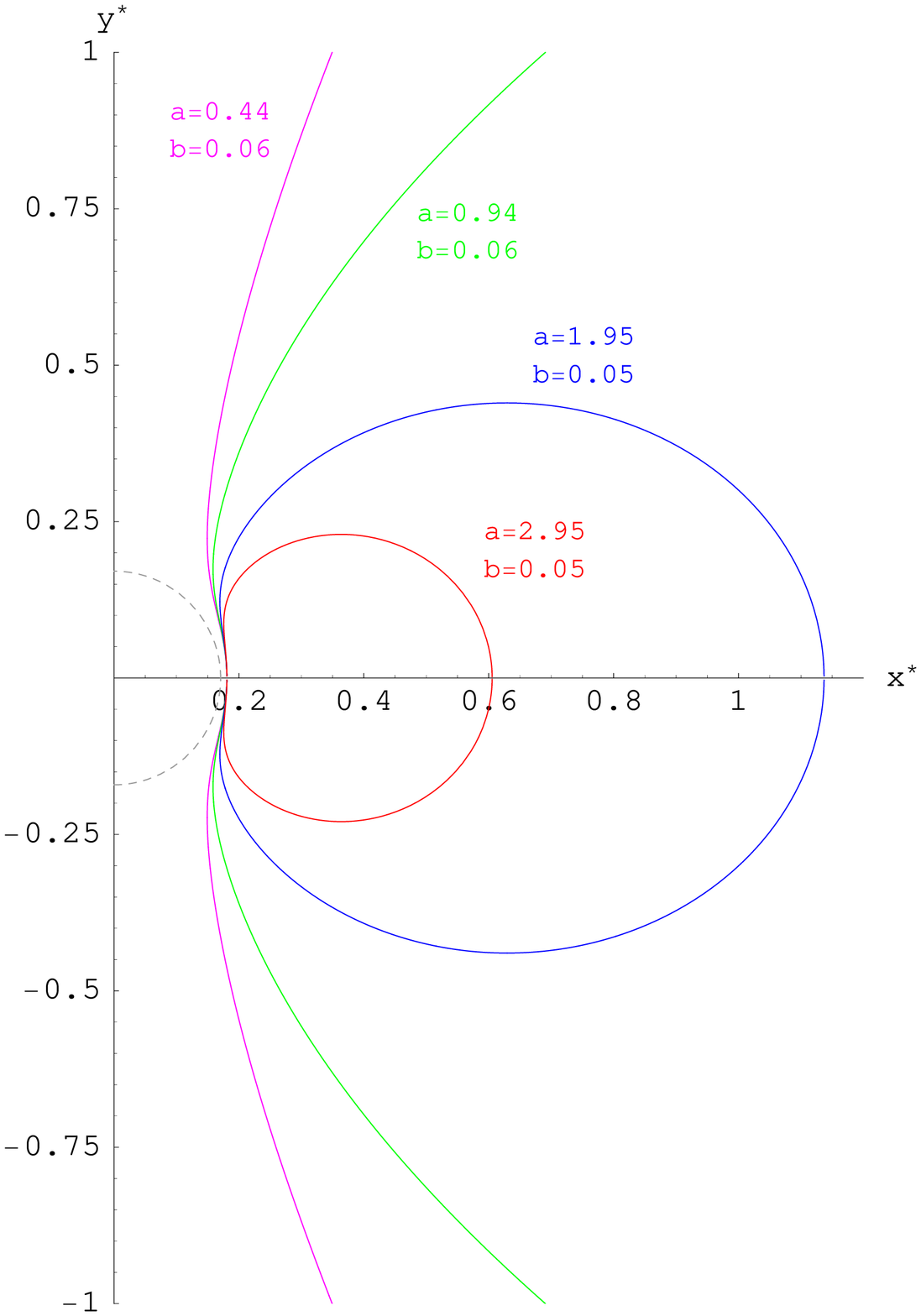}\hspace*{1cm}
\caption{A mixture of brane trajectories in a 5-dimensional
Schwarzschild--anti de Sitter background of fixed parameters $k=1$
and $\mu=0.03$.}
\label{fig:mixture}
}
%%%%%%%%%%%%%%%%%%%%%%%%%%%%%%%%%%%%%%

\subsection{Subcritical branes}

This family of branes with $a+b<1$ are largely similar to critical branes.
They correspond to open trajectories that asymptote the adS boundary,
although at nonzero $\chi$ in this case. 
The same bound as before, i.e.\ whether
$|\cos \chi| \simeq |a e^{\tilde r_+} + b e^{-\tilde r_+}| \leq 1$,
will determine whether the brane terminates on the event horizon
or remain on the RHS of it. As the brane trajectories in this case
look similar, apart from the angle of approach from the adS boundary,
to the ones presented in figure \ref{fig:crit}(a) for critical branes,
we refrain from presenting another graph here. The energy density
and pressure profile in this case is again similar to the one found
for critical branes.
Once again, for a large family of parameters $a$ and $b$, solutions
with a positive energy excess at the center of the brane may be
easily found.

One special subcritical trajectory found in the pure adS case was the 
Karch Randall trajectory, $a+b=0$. We can extend this to Sch-adS 
obtaining
\be
\cos\chi = 2a \sinh {\tilde r}
\ee
however, since $a>0$ for a positive energy trajectory, this has
$(\cos\chi)'>0$, and hence the energy density is always increasing
with $r$. Thus, whether or not these trajectories terminate on the horizon,
they always correspond to energy deficits on the brane, and hence
negative mass sources from the point of view of a brane observer.

Finally, one other special trajectory that emerged from the previous
section was the Schwarzschild trajectory which was nonsingular on
the horizon, intersecting it perpendicularly and extending to the
origin. This corresponds to $ae^{{\tilde r}_+} = b e^{-{\tilde r}_+}$
(note that the condition $A=B$ for Schwarzschild was for an ${\tilde r}$
coordinate zeroed at the horizon). Extending this concept to Sch-adS gives
\be
\cos\chi = 2ae^{{\tilde r}_+} \cosh ( {\tilde r} - {\tilde r}_+ )
\ee
These trajectories can be super- sub- or precisely critical, depending on
the magnitude of $a$, however, for all of these trajectories $(\cos\chi)'>0$,
hence they correspond to energy deficits on the brane.

The brane trajectories found in this section and depicted in figures
\ref{fig:supercrit} and \ref{fig:crit} have obvious similarities with
the ones presented in the previous section.
As expected, the brane trajectories in a bulk containing both a
mass and a negative cosmological constant are hybrid constructions,
and exhibit features and characteristics that appeared either
in the case of an adS or a Schwarzschild background. As an example,
in figure \ref{fig:mixture}, we present a mixture of supercritical,
critical and subcritical branes embedded in a Sch-adS bulk, that 
clearly resembles the one presented in figure \ref{fig:ads1}.
The parameters $a$ and $b$ in this particular case have been 
chosen so that the branes remain on the RHS of the horizon, and
apart from the brane bending to avoid the horizon, these trajectories
have otherwise similar characteristics with the ones in figure \ref{fig:ads1}. 
Finally, we would like to note here that the study of Sch-adS 
backgrounds with larger mass parameter $\mu$ has led to similar
families of trajectories. As mentioned earlier, as the value of
$\mu$ gets bigger, the range of the $\tilde r$-coordinate becomes
shorter, and an increased accuracy is necessary in our numerical
analysis in order to produce the corresponding trajectories.
Apart from being numerically more sensitive, the study of 
large mass Sch-adS backgrounds yields the same characteristics
for the allowed brane trajectories.

%%%%%%%%%%%%%%%%%%%%%%%%%%%%%%%%%%%%%%

\section{The Time-Dependent Brane}\label{sec:tdep}

Having analysed the static brane trajectories in a spherically symmetric
background, finding brane black hole and stellar solutions, we now 
comment on the time dependent case. The full problem is beyond
the scope of this paper, however, we will make some remarks here, and
explore the various issues involved in finding a time-dependent brane
black hole trajectory, such as might be appropriate to black hole 
evaporation or recoil.

\subsection{Exact branes}

We start by reviewing the argument that there is no time-dependent
trajectory which corresponds to a pure vacuum brane embedded in a
black-hole bulk background.  Setting $v=-1$ in the brane equations 
(\ref{eq:IsraelTTalpha})-(\ref{eq:ConsistenceRalpha})
results in considerable simplifications. Once again, the energy 
density
\begin{eqnarray}
\rho = \frac{\rho_0}{r^2}\,\left[U(\alpha^\prime r - \alpha) +
\alpha\right]\,, \label{eq:cosmorho}
\end{eqnarray}
(which in principle can be a $(\tau,r)$-dependent quantity) solves
(\ref{eq:ConsistenceTalpha}) and (\ref{eq:ConsistenceRalpha}). 
In addition, (\ref{eq:IsraelTRalpha}) is immediately integrable
yielding
%%%%%%%%%%%
\begin{equation}
\alpha(\tau, r)= f(\tau)\,\sqrt{U(r)} + g(r)\,, \label{sol-time}
\end{equation}
%%%%%%%%%%%%
where $f(\tau)$ and $g(r)$ are, at the moment, arbitrary functions. 
If we substitute the above form of $\alpha(\tau, r)$ into
(\ref{eq:IsraelRRalpha}), we then find
%%%%%%%%%%%%
\begin{eqnarray}
\sqrt{U}\left(\frac{1}{2}\,U^{\prime\prime} r^2 - U^\prime r
+ U-1\right)f(\tau) + Ur^2g^{\prime\prime} + (g^\prime r - g)
\left(\frac{U^\prime r}{2} - U\right) - g =0.\,
\label{RR-again}
\end{eqnarray}
The only way for this equation to be satisfied, for all $\tau$,
with $f(\tau) \neq 0$, is for the factor preceding $f(\tau)$ to
be equal to zero:
\begin{eqnarray}
\frac{1}{2}\,U^{\prime\prime} r^2 - U^\prime r + U-1 =0\,,
\end{eqnarray}
which has as a solution
\begin{eqnarray}
U(r) = 1 + Cr + Dr^2\,, \label{U-time}
\end{eqnarray}
for constants $C$ and $D$. Thus a brane
with equation of state $w=-1$ can only be embedded in 
a bulk with the above solution for the metric function.
However, computing the energy momentum tensor of the bulk spacetime,
and demanding that it be isotropic leads to $C=0$, and hence a 
constant curvature spacetime with $D=-\Lambda/6$. Clearly
adS spacetime satisfies this, with $D=k^2$. Here the
pure tension brane trajectory takes the form
%%%%%%%%%%%%%%
\be
\alpha(\tau, r)=\frac{1}{k}\,\sqrt{1+k^2 r^2}\,\cos (k \tau)
-\frac{1}{k}\,,
\ee
%%%%%%%%%%%%%
with energy, as expected,
%%%%%%%%%%%%
\be
\rho=\rho_{RS}=\frac{6k}{\kappa_5}\,.
\label{rho-RS}
\ee
%%%%%%%%%%%%%%%
This is the Randall-Sundrum brane in global coordinates.
To see this, use the transformation between global adS and RS 
coordinates:
\bea
ku &=& \left [ \sqrt{1 + k^2 r^2} \cos k\tau
- kr \cos \chi \right ]^{-1} \label{utoGL} \\[1mm]
kt &=& \left ( ku \right ) \sqrt{1+k^2r^2} \ \sin k\tau \label{ttoGL} \\[1mm]
k|{\bf x}| &=& \left ( ku \right) kr \sin \chi\,,\label{xtoGL}
\eea
in which the RS metric is:
\be
ds^2 = \frac{1}{k^2u^2} \left [ -dt^2 + du^2 + d{\bf x}^2 \right ]\,.
\ee
The wall trajectory in global coordinates
is in fact oscillatory, starting off at the adS boundary, moving in to
the origin when it closes off the whole adS boundary, then moving
back again. The RS wall is 
oscillatory because the spherical coordinates are the
universal covering space of adS, and so the `wall' is actually an
infinite family of walls, each in the local patch covered by the
horospherical coordinates. 
 
The key feature of this wall trajectory is that it satisfies
the Israel junction conditions \cite{Israel} for a `brane' 
energy-momentum tensor which is proportional to the induced metric
on the brane. A natural generalization of this situation would be 
the consistent embedding of a brane, with a general energy-momentum
tensor, in a bulk background that describes a regular black hole in
either flat or curved spacetime. 
 
\subsection{Branes with matter}

Let us now explore simplistically what happens if we modify the global
adS metric to Sch-adS.
Since $r=0$ is a geodesic of the spherical
adS spacetime, the image of $r=0$ in the Randall-Sundrum spacetime,
which is a hyperbola, will be a geodesic in the RS spacetime.
Therefore, if we put a black hole at $r=0$, it should look
like a particle in the RS spacetime, at least to a first approximation.
We should note here that this philosophy is similar to that of 
\cite{Galfard}, who considered slices of Sch-adS
which would satisfy the DEC. However, note that those slices
were static (and asymptotically adS), crucially here we are 
allowing time dependence in order to get an asymptotically flat brane,
as well as the interpretation of black hole evaporation or recoil

As we have just shown, it is not possible to have a vacuum brane
embedded in anything other than pure adS spacetime, therefore
finding a full solution to a time-dependent braneworld black hole 
would require embedding a time-dependent axisymmetric brane at
the very least in an axisymmetric bulk, since it was
proven in \cite{BCG} that a bulk with SO(4) symmetry must necessarily
be Sch-adS. Thus, the exact solution for the vacuum brane is
an extremely involved problem which is probably only tractable numerically. 
What we will now do is to relax the assumption that $T_{\mu\nu} \propto
h_{\mu\nu}$ for the brane, and explore the effect of the bulk black hole
on the {\it time-dependent} brane, determining the effect of brane bending
and varying the brane position and black hole mass on the 
brane energy-momentum.

From (\ref{utoGL}), we see that a
flat brane sitting at $ku = \epsilon$ in the RS picture
satisfies 
\be
\sqrt{1+k^2r^2} \cos k\tau - kr\cos\chi = (ku)^{-1} = \frac{1}{\epsilon}\,.
\label{RSwallgenu}
\ee
Therefore as a starting point, we keep this same trajectory and
calculate the energy-momentum.  As we have already discussed,
there is no trajectory that will have a pure brane energy momentum,
$T_{\mu\nu} = -(6k/\kappa_5)\,h_{\mu\nu}$, however, as a black hole
formed on the brane must be produced by the collapse of matter, 
it is not clear whether we should expect a pure
brane energy momentum solution; rather,
a solution corresponding to the collapse of matter on the brane is
perhaps more physically realistic.

The aim of this subsection is to understand how
the bulk black hole acts on the brane energy momentum, producing
a spherically symmetric brane matter source.
By varying the black hole mass, the distance of closest approach of
the black hole to the brane, and modifying the brane trajectory, 
we can study the effect of these various factors on the brane 
energy momentum.
Since we expect the full picture to involve all three of
these effects in some combination, this process of isolating the effect
of each will allow us to gain insight into the nature of
the brane black hole.
Unlike the previous sections however, here we are not imposing a particular
type of matter, such as an isotropic fluid, but determining
whether time-dependent brane trajectories exist which have 
sensible matter on the brane. In particular, we will be looking
for matter which obeys the weak energy condition (WEC),
namely that the energy density increases at the center of collapse.

Recall that the energy momentum of a surface slicing the Sch-adS
spacetime is given by the Israel junction conditions as:
\be
T_{\mu \nu}= \frac{2}{\kappa_5}\left ( K_{\mu \nu} - K h_{\mu \nu}\right ) 
= -\frac{6k}{\kappa_5}\,h_{\mu \nu} +  t_{\mu \nu}\,,
\ee
where we have taken the critical RS brane tension $6k/\kappa_5$, 
and $ t_{\mu \nu}$ is (hopefully) a small perturbation in the 
energy-momentum. We can compute this geometric energy momentum
from (\ref{TT}-\ref{ThTh}), and since we want to compare this to a pure
critical RS brane $T_{\mu\nu}= -(6k/\kappa_5)\,h_{\mu\nu}$ 
we will plot the ratio (in an obvious abuse of notation):
\be
e_{\mu\nu} = 
\frac{(K_{\mu\nu} - K h_{\mu\nu})}{(-h_{\mu\nu})} = 
\frac{T_{\mu \nu}}{(-2h_{\mu \nu})}= 3 +
\frac{t_{\mu \nu}}{(-2h_{\mu \nu})}\,.
\ee
In the above, we have set for simplicity $\kappa_5=1$.
Clearly, since the trajectory is time dependent, this ratio will also be
time dependent, however, we expect that the largest effect of the bulk
black hole will be represented by the $t=0$ slice of the braneworld
-- the point of closest proximity --
we therefore plot the energy momentum at $t=0$. This has the 
simplifying effect of making ${\dot\chi}=0$ in the expressions for 
the extrinsic curvature, and gives a local static frame for the energy
momentum tensor. 
\FIGURE{
\includegraphics[width=6cm]{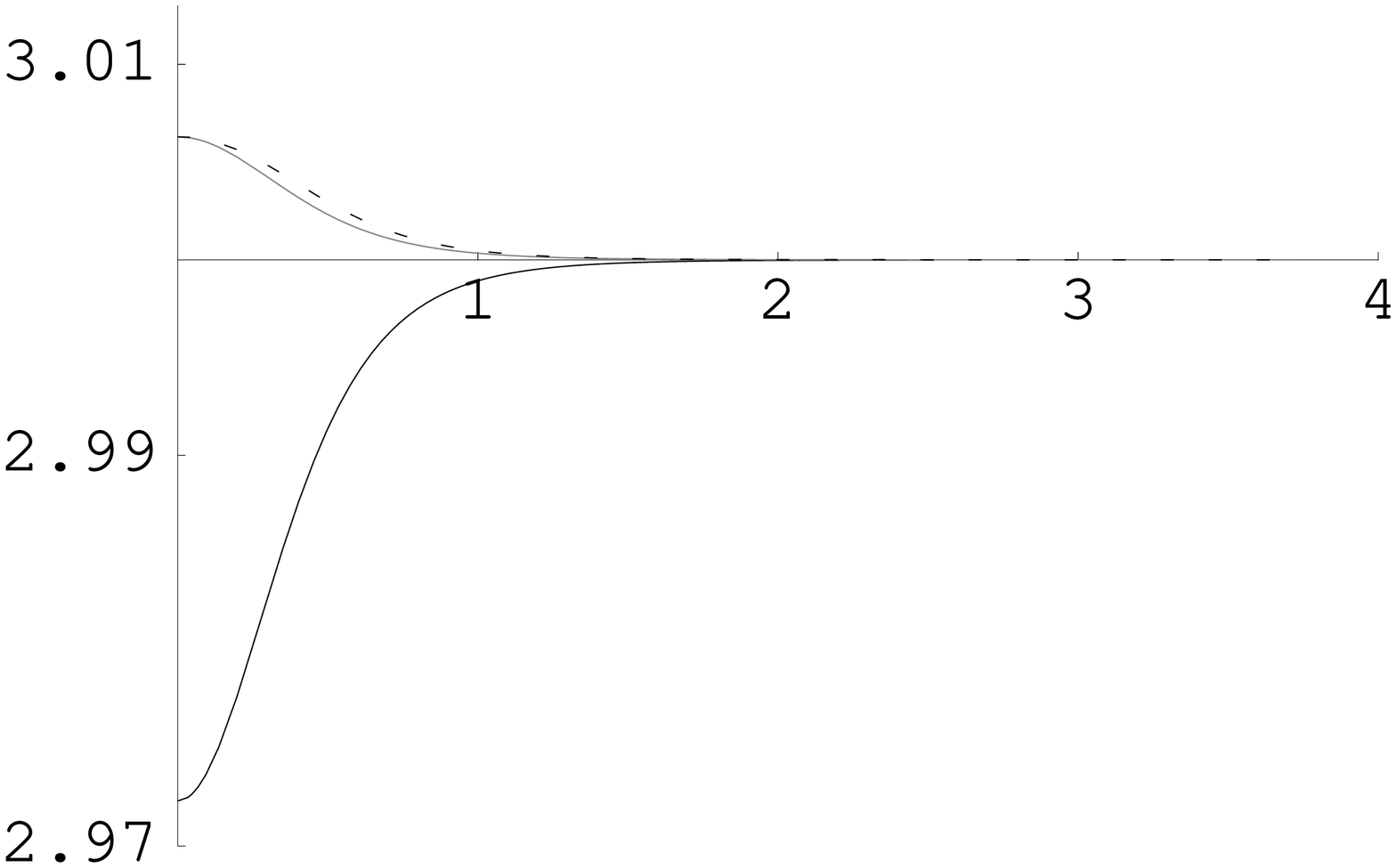}
\caption{The energy-momentum of an RS brane with a black hole in the
bulk as a function of the brane radial coordinate $r\sin\chi$. 
The energy is given by the black line, the radial pressure by the grey
line and the azimuthal pressure by the dotted line.}
\label{fig:RSbrane}
}

Figure \ref{fig:RSbrane} shows the effect of the bulk black hole
on the energy momentum tensor of the brane. We have picked a small bulk
black hole ($\mu=0.1$), and the RS brane sits at $ku = 0.3$. The energy
is the solid black line, the radial tension the grey line, and the
angular tension the dotted line.
The black hole causes the energy of the brane to decrease from its critical
value, whereas both the radial and azimuthal tension increase. 
Clearly, therefore, from the point of view of differential energy-momentum,
this brane does not satisfy the WEC, although the overall energy momentum
does satisfy the WEC. The brane however fails to satisfy the 
dominant energy condition (DEC). Unsurprisingly, moving the RS brane
towards the black hole increases the overall effect on the energy and tension.
Increasing the mass of the black hole in the interior changes this 
picture surprisingly little, provided we simultaneously move the RS brane
towards the boundary.
%%%%%%%%%%%%%%%%%%%%%%%%%%%%%%%%%%%%%%
\FIGURE{
\includegraphics[width=5cm]{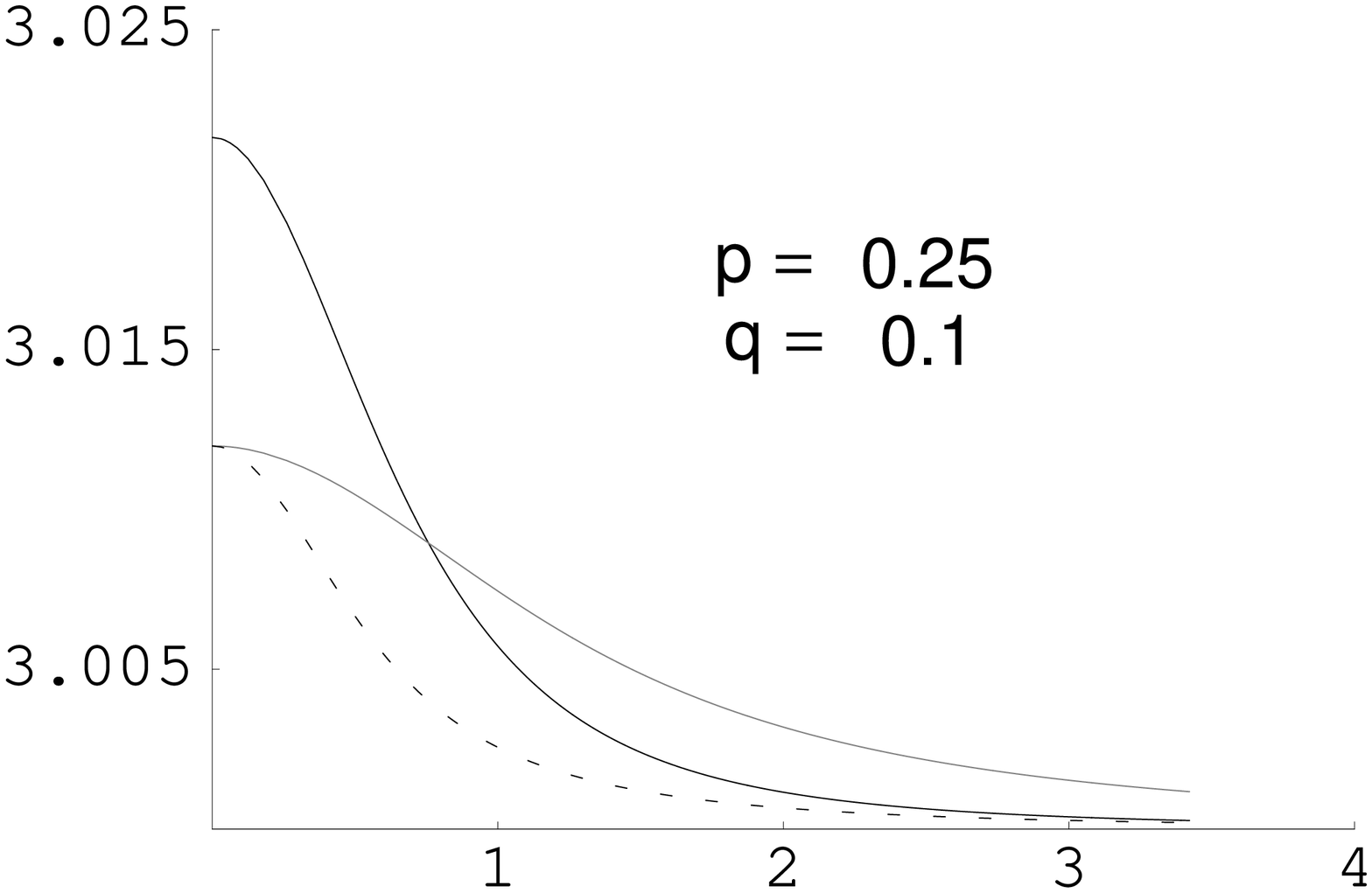}\nobreak\hskip3mm\nobreak
\includegraphics[width=5cm]{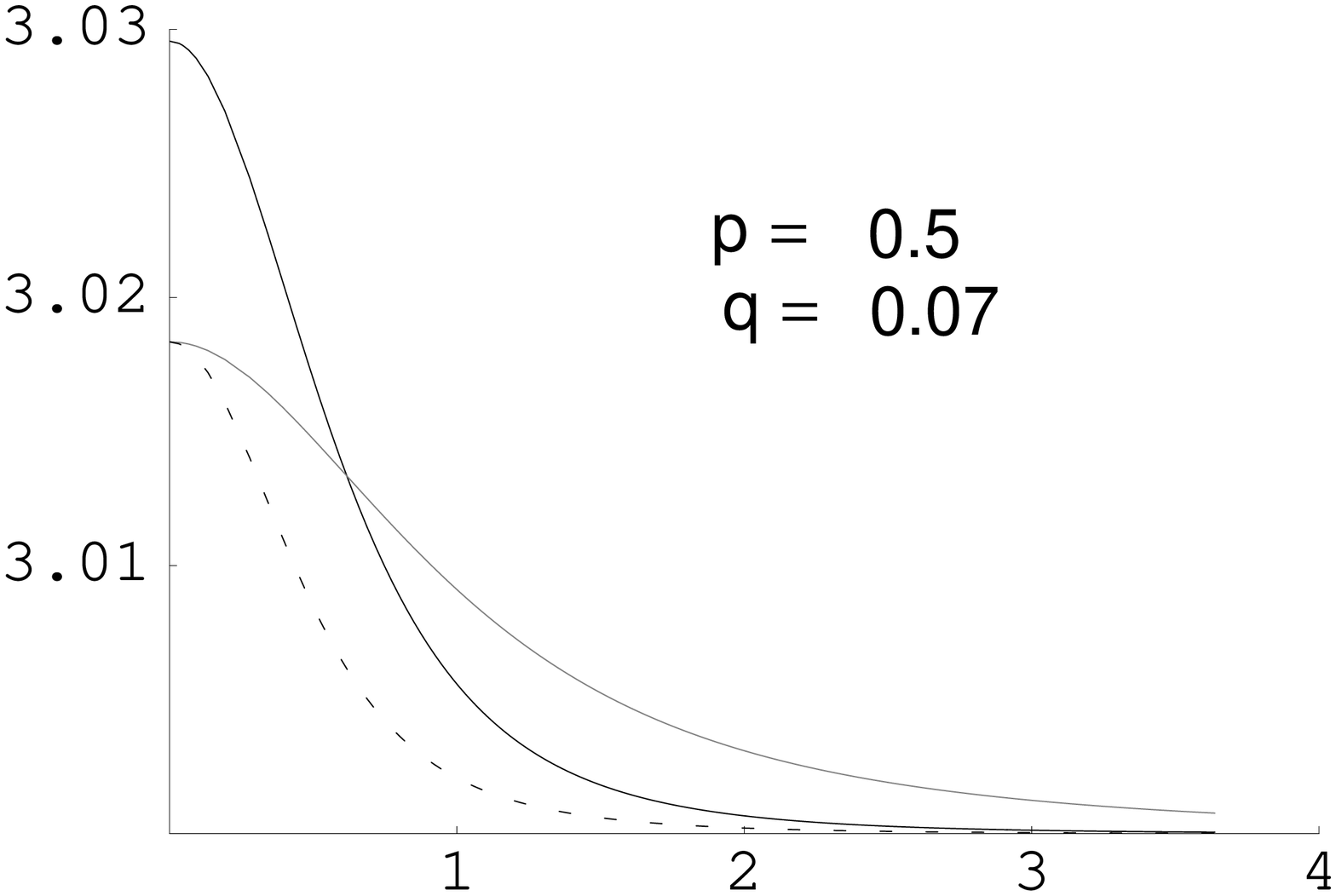}\nobreak\hskip3mm\nobreak
\includegraphics[width=5cm]{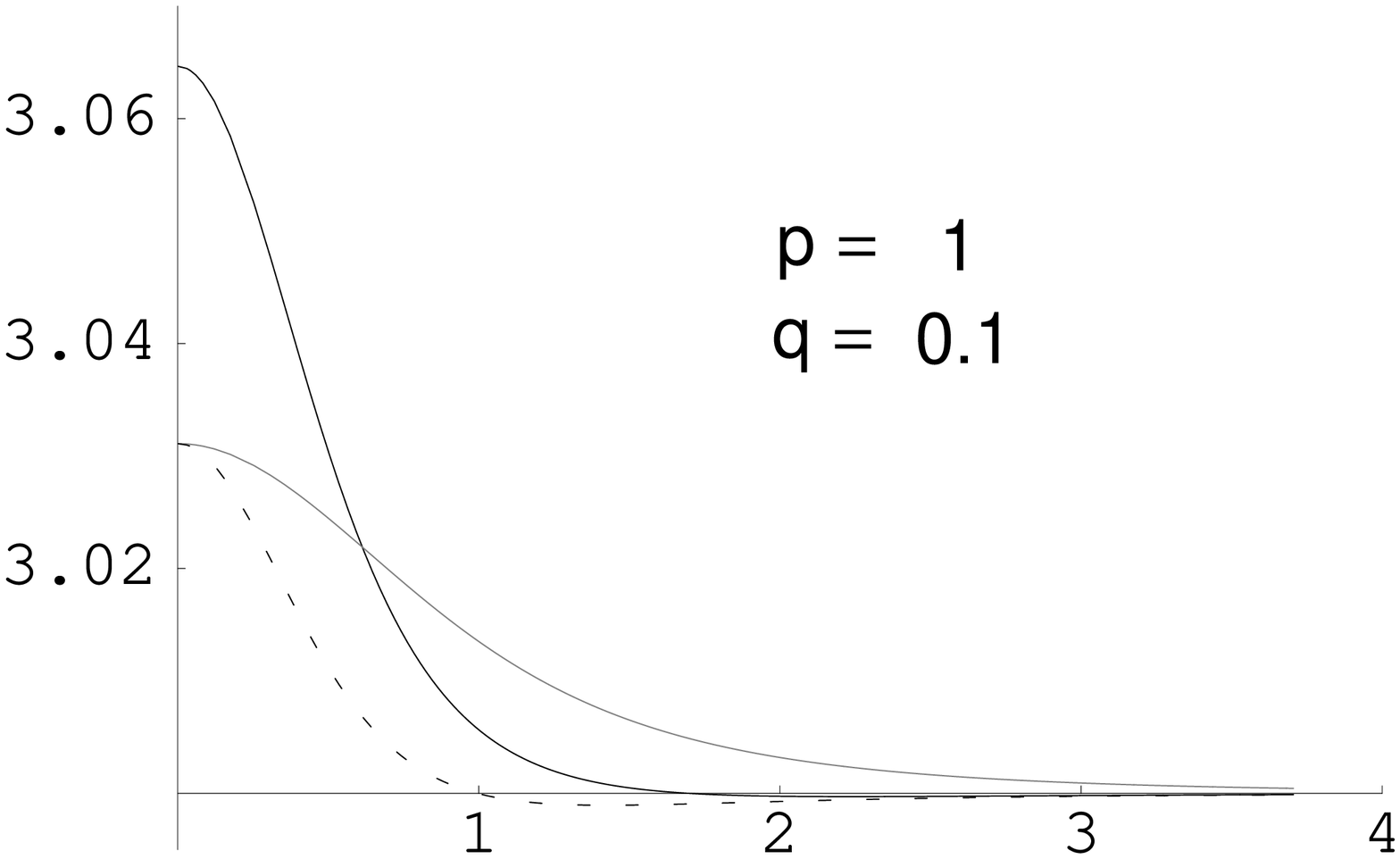}
\caption{A selection of plots of brane energy-momentum with brane bending
included for a range of amplitudes and powers of $r$.}
\label{fig:bend}
}
%%%%%%%%%%%%%%%%%%%%%%%%%%%%%%%%%%%%%%

Clearly therefore, maintaining the RS slicing of the global spacetime
in the presence of a black hole results in brane energy momenta which are
not physically sensible, they satisfy neither brane WEC nor DEC. However,
this was not unexpected, as we have not modified the RS trajectory,
and the main feature of our static brane solutions was that they
responded to the bulk black hole by bending. Indeed, in a
definitive brane gravity paper, \cite{GT}, Garriga and Tanaka
showed that a crucial part of obtaining four dimensional
{\it Einstein} gravity (i.e.\ with the correct tensor structure)
was what could be interpreted as a brane bending term. Essentially,
the effect of matter on the brane was to ``shift'' the brane with
respect to the acceleration horizon in the bulk.
Clearly then, if a black hole forms on the brane, we would expect
the brane to respond to this matter by bending. The $1/r$ brane bending
term derived in \cite{GT} was of course in the static limit, therefore
we cannot simply read off what the non-static term should be,
therefore we try a variety of modifications to the position of
the brane to explore their effect.

A shift in the position of the brane corresponds to $ku \to 1 +
k\delta u$, and we will make the simplifying assumption that
$k\delta u = -\delta(r)$, trying out a range of functions $\delta$:
\be
\delta(r) = \frac{q}{r^p}\,.
\ee
What this does is to change the function $\chi$:
\be
\cos \chi(\tau,r)=\frac{1}{r} \left (\sqrt{1+ r^2}\
\cos\tau - \frac{1}{u} \right ) \simeq \frac{1}{r}
\left (\sqrt{1+ r^2}\ \cos\tau - \frac{1}{1+ \delta(r)} \right )\,,
\ee
where we can use the transformation between RS and global 
coordinates to try out a range of power laws in $\delta(r)$
which reduce to sensible brane bendings in the RS picture. 

We have looked at the effect of bending the brane both towards and
away from the black hole for a variety of 
$\delta(r)$. In all cases, bending the brane towards the black hole 
worsens the energy deficit on the brane, and does not help with the
DEC. However, bending the brane {\it away} from the black hole,
as occurs in the static brane trajectories, can
remove the energy deficit, and indeed restore
the DEC near the centre of the brane. This indicates that a true brane
trajectory will try to avoid the black hole by bending away from it.

Figure \ref{fig:bend} shows the effect of the brane bending on the
energy-momentum. Bending the brane by a power $p>1$ introduces an 
energy deficit at some radius, therefore we only show those functions
which maintain the differential brane WEC. The GT brane bending of $1/x$
corresponds approximately to $p=1/2$.

Finally, we point out that unlike the static trajectories, in these
time dependent cases the black hole is in the bulk spacetime, hence 
these are candidate branes for a black hole having recoiled into the bulk.

\subsection{The interaction of black holes and branes}

The main motivating factors for obtaining a time-dependent 
braneworld black hole are to gain insight into the 
back-reaction of Hawking radiation on
a quantum corrected four-dimensional black hole, and to 
understand the process of black hole recoil from a braneworld. 
In the former case we would want a $\mathbb{Z}_2$-symmetric 
solution, while the latter case does not necessarily require this.
Since we have been working within the $\mathbb{Z}_2$-symmetric 
setup, we will only consider this here. 

The idea is that the
time-dependent process will be some perturbed version of a
time-dependent brane trajectory in five-dimensional
Sch-adS spacetime. By allowing the brane to intersect the 
bulk black hole horizon, this would appear to describe 
black hole formation and  evaporation via transport of a 
bulk black hole to the brane, and subsequent departure back into the
bulk. 
When the brane hits the black hole, we might expect
some part of it will be captured
by the black hole, and will therefore remain behind the event  horizon even
when the black hole has left the brane, effectively having been chopped
off from the rest of the brane. This feature is seen in the probe brane
calculations of \cite{PrBr}, and we expect this to hold in the case of
a fully gravitating brane. In support of this, we can appeal to 
the case of a cosmic string interacting with a black hole, 
where early work indicated that strings would be captured, and via
self-intersection would leave some part behind in the black hole \cite{CSBH},
and idealized gravitational calculations show explicitly how this ties in
with the thermodynamic process of string capture and black hole entropy
\cite{ABEGK}. Just as in the cosmic string case, this capture of the brane by
the black hole will turn out to be important in establishing the 
thermodynamic viability of the black hole recoil process.

The recoil of the black hole from the brane is an important indicator of the
possible evaporation process. Since the energy-momentum on the brane violates
the DEC, it might be thought that this is an indicator that
the black hole cannot leave the brane, and indeed a naive entropy argument
supports this view. The argument runs as follows: A black hole of mass
$M$ on the brane has entropy $\sim M^{3/2}$, however, if it recoils into
the bulk, it must recoil as two black holes of mass $M/2$, hence total
entropy $2(M/2)^{3/2}=M^{3/2}/\sqrt{2}$. Thus black hole recoil is 
entropically forbidden. However, there are some interesting possible
corrections to this simple argument, 
that can change this equation rather substantially. First, 
at the calculational level, the fact that
entropy is proportional to horizon area/volume, which for
Sch-adS is not simply related to the mass. 
Next, it does not take into account
the effect of the brane bending. Since the brane is bent, if it does
intersect the black hole horizon, it will not do so at the
equator, but somewhat ``south'', i.e.\ for $\chi_0>\pi/2$.
For the black hole intersecting the horizon at 
$\chi_0 $, the actual area contributing to the entropy of
the intermediate state is
\be
{\cal A}_{int} = 4\pi f(\chi_0)
\left [ \frac{\sqrt{1+2\pi k^2\mu/f(\chi_0)}-1}{2k^2}\right]^{3/2}\,,
\ee
where $f(\chi) = \chi/2 - (\sin 2\chi)/4$.

Finally, in the process of the black hole intersecting the brane, we
would expect some part of the brane to be excised and captured by the
black hole, thus increasing its mass. Estimating this as 
\be
\delta M = \frac{6k}{\kappa_5} \frac{4\pi}{3} (r_+\sin\chi_0)^3\,,
\ee
we see that the area of the final state is in fact
\be
{\cal A}_f \simeq {2\pi^2} 
\left [ \frac{\sqrt{1+4k^2\mu + 32k^3(r_+\sin\chi_0)^3/3\pi}-1}{2k^2}
\right]^{3/2}\,.
\ee
The above becomes larger than the intermediate area at $k^2\mu \simeq 0.01$
for $\chi_0\leq \pi/4$, and even ignoring the brane bending effect
by setting $\chi_0=\pi/2$, we still find that black hole recoil is
preferred at $k^2\mu \geq 0.07$, or at a (four-dimensional) 
Schwarzschild radius
\be
2G_Nm \simeq l_{\,adS}/10
\ee
where $l_{\,adS}$ is the adS length scale.

It is important to note that these arguments use the
standard entropy of the isolated Sch-adS black hole. In
other words, they assume a static solution with an event horizon
at $r_+$. Clearly in the time-dependent spacetime there is some
question about whether this approximation is valid, indeed, even
ignoring branes, we see that a black hole with mass greater than
about $3$ (in adS units) has lower entropy than two black holes of mass
$3/2$. Clearly therefore, entropy arguments should be used with caution,
nonetheless, for small black holes, where we might expect them to
be more reliable, taking into account brane bending
and fragmentation shows that it is by no means entropically preferred
for a black hole to stick to the brane.

To sum up: taking time dependent brane trajectories in a Sch-adS 
background give brane trajectories with additional matter on the brane.
The brane prefers to bend away from the black hole, and in such a case
the matter satisfies the DEC near the core of collapse, and the WEC
at all times. For very small black holes ($\mu k^2 \leq 0.01$) our
results suggest that the bending of the brane may cause it to 
avoid the black hole altogether. We cannot use our approximation to
draw conclusions for large black holes, as it presupposed the black hole
was a small perturbation to the RS spacetime. The entropy argument for
black hole recoil supports the notion that part of the brane will be
captured by the black hole, however, we have not attempted to model
this gravitationally with time dependent closed branes at this
point. Clearly any more detailed calculation would require modification
of the bulk solution.

%%%%%%%%%%%%%%%%%%%%%%%%%%%%%%%%%%%%%%%%%%%%%%%%%%%%%%

\section{Conclusions}

In this work, we have analyzed spherically symmetric brane solutions
in a known bulk spacetime with the aim of finding a consistent black
hole solution for the brane. We found that the problem of a static
braneworld slicing a known spherically symmetric bulk was completely
integrable, with the solution being given in terms of an implicit
function of the bulk radial variable. Thus, we have found all
possible complete
brane and bulk solutions for a brane with a perfect fluid matter
source living on it -- in other words complete brane TOV solutions.
These solutions have the interpretation of braneworld
stars, and correspond to static slicings of a Sch-adS bulk spacetime,
with the bulk solution corresponding to the part of the Sch-adS 
spacetime {\it not} containing the event horizon of the black hole.
Thus our solutions are completely nonsingular. 
We have also found solutions in which the event horizon of a bulk
black hole impinges upon the brane, but these typically have 
divergent pressure on the brane, reminiscent of the singularity in
the TOV system when we try to solve for too big or compact a star.

All of our solutions contain excess pressure at large radii, this seems
to be a feature of the slicing of the pure adS bulk, and it is related
to the fact that the Randall-Sundrum solution, a pure Minkowski brane,
is in fact not a static slicing of adS in global coordinates. The
only possibility for having a well-behaved asymptopia is to have a
subcritical Karch-Randall brane. These however, cannot be extended into
positive mass sources. We have therefore been unable to find a solution
which has all the features we would desire in a braneworld star, however,
we have made crucial progress by first demonstrating how to find
exact and complete solutions to the brane TOV problem, as well as
classifying these according to their energy-momentum.

Probe brane calculations of the interaction of a wall with a black hole
indicate the possibility of brane excision, that is, that as the black hole
leaves the brane, the brane is distorted so much that it self-intersects
and part of the brane is excised, falling into the black hole, with the
remainder moving away towards infinity.
Among our various solutions are closed bubbles as well as open branes,
and it is tempting to try to model this process using a quasi-static
approach -- taking a sequence of the static solutions we have found as
approximate solutions (such as in \cite{CritVe}). Unfortunately however, 
positivity of energy requires that the interior of the bubbles be kept,
and for the branes extending to infinity, the bulk does not contain
a black hole, therefore these solutions are not suitable for such
an approximation. If we wished
to keep the exterior of the bubble and the black hole inside the bulk
spacetime, we would need negative energy branes. It would appear
that time-dependence is key to finding consistent solutions, as
in those cases, the black hole is actually retained in the bulk.

We have also explored the time dependent brane trajectories, using
the RS brane as a starting point, to try to model the process of
gravitational collapse, and to explore the issues involved in black
hole evaporation. We found that the effect of a bulk black hole on 
the RS brane was to induce a negative energy source, however, by 
bending the RS brane by a small amount, we could restore the brane-DEC,
although these solutions had anisotropic pressure. Provided we allow
matter on the brane, we can form trajectories which now have black holes
in the bulk, and which can intersect with the brane, although in this
case the DEC is violated.

Finally, an important point to note is that in all our results, we have
made the simplifying assumption of $\mathbb{Z}_2$-symmetry around the brane. 
The RS model is $\mathbb{Z}_2$-symmetric, and many of the investigations
into gravitational braneworld solutions are also $\mathbb{Z}_2$-symmetric.
However, it is obviously important to check and explore if any of
our conclusions change significantly if we drop this restriction. In 
particular, for black hole recoil off the brane, we would expect the
black hole to recoil on one side only of the brane, and hence for
$\mathbb{Z}_2$-symmetry to be broken. It may be that many of the restrictions
we have found with our solutions can be evaded if we remove
$\mathbb{Z}_2$-symmetry. This is currently under investigation.

\acknowledgments 
We would like to thank Roberto Emparan for useful discussions.
S.C.\ and P.K.\ are supported by PPARC Grants PPA/S/S/2004/3815 and
PPA/A/S/2002/00350 respectively, and B.M.\ is supported by EPSRC.

\end{document}